\documentclass[twocolumn]{aastex62}
\usepackage{CJK}
\usepackage{xcolor}

\newcommand{\dm}{$\Delta m_{15} (B)$}
\newcommand{\sbv}{$s_{\mathrm{BV}}$}
\newcommand{\um}{$\mu$m}
\newcommand{\hei}{\ion{He}{1}}
\newcommand{\ci}{\ion{C}{1}}
\newcommand{\cii}{\ion{C}{2}}

\newcommand{\siii}{\ion{Si}{2}}

\newcommand{\mgii}{\ion{Mg}{2}}
\newcommand{\caii}{\ion{Ca}{2}}

\newcommand{\lam}{$\lambda$}
\newcommand{\nirad}{$^{56}$Ni}
\newcommand{\nista}{$^{58}$Ni}
\newcommand{\corad}{$^{56}$Co}
\newcommand{\nifor}{[\ion{Ni}{2}]}
\newcommand{\fefor}{[\ion{Fe}{2}]}
\newcommand{\halpha}{H${\alpha}$}
\newcommand{\palpha}{Pa${\alpha}$}
\newcommand{\pbeta}{Pa${\beta}$}
\newcommand{\pgamma}{Pa${\gamma}$}
\newcommand{\sneia}{SNe\,Ia}
\newcommand{\snia}{SN\,Ia}

\newcommand{\sneincp}{21}
\newcommand{\sneinc}{125}
\newcommand{\sneinp}{90}
\newcommand{\snetot}{157}
\newcommand{\nirspecinp}{451}
\newcommand{\nirspectot}{661}

\shorttitle{NIR spectroscopy of supernovae}
\shortauthors{Hsiao et al.}

\begin{document}
\begin{CJK*}{UTF8}{bsmi}

\title{Carnegie Supernova Project-II: The Near-infrared Spectroscopy
  Program\footnote{This paper includes data gathered with the 6.5-m
    Magellan telescopes at Las Campanas Observatory, Chile.}}

\author[0000-0003-1039-2928]{E. Y. Hsiao (蕭亦麒)}
\affil{Department of Physics, Florida State University, 77 Chieftan Way, Tallahassee, FL 32306, USA}

\author[0000-0003-2734-0796]{M. M. Phillips}
\affil{Carnegie Observatories, Las Campanas Observatory, Colina El Pino, Casilla 601, Chile}

\author{G. H. Marion}
\affil{Department of Astronomy, University of Texas, 1 University Station C1400, Austin, TX 78712, USA}

\author[0000-0002-1966-3942]{R. P. Kirshner}
\affil{Harvard-Smithsonian Center for Astrophysics, 60 Garden Street, Cambridge, MA 02138, USA}
\affil{Gordon and Betty Moore Foundation, 1661 Page Mill Road, Palo Alto, CA 94304, USA}

\author[0000-0003-2535-3091]{N. Morrell}
\affil{Carnegie Observatories, Las Campanas Observatory, Colina El Pino, Casilla 601, Chile}

\author[0000-0003-4102-380X]{D. J. Sand}
\affiliation{Department of Astronomy and Steward Observatory, University of Arizona, 933 N Cherry Avenue, Tucson, AZ 85719, USA}

\author[0000-0003-4625-6629]{C. R. Burns}
\affil{Observatories of the Carnegie Institution for Science, 813 Santa Barbara St, Pasadena, CA 91101, USA}

\author[0000-0001-6293-9062]{C. Contreras}
\affil{Carnegie Observatories, Las Campanas Observatory, Colina El Pino, Casilla 601, Chile}

\author[0000-0002-4338-6586]{P. Hoeflich}
\affil{Department of Physics, Florida State University, 77 Chieftan Way, Tallahassee, FL 32306, USA}

\author[0000-0002-5571-1833]{M. D. Stritzinger}
\affil{Department of Physics and Astronomy, Aarhus University, Ny Munkegade, DK-8000 Aarhus C, Denmark}

\author[0000-0001-8818-0795]{S. Valenti}
\affil{Department of Physics, University of California, Davis, CA 95616, USA}

%alphabetical

\author[0000-0003-0227-3451]{J. P. Anderson}
\affil{European Southern Observatory, Alonso de C\'ordova 3107, Casilla 19, Santiago, Chile}

\author[0000-0002-5221-7557]{C. Ashall}
\affil{Department of Physics, Florida State University, 77 Chieftan Way, Tallahassee, FL 32306, USA}

\author[0000-0003-0424-8719]{C. Baltay}
\affil{Physics Department, Yale University, 217 Prospect Street, New Haven, CT 06511, USA}

\author[0000-0001-5393-1608]{E. Baron}
\affil{Homer L. Dodge Department of Physics and Astronomy, 440 West Brooks Street, Room 100, Norman, OK 73019, USA}

\author{D. P. K. Banerjee}
\affil{Astronomy and Astrophysics Division, Physical Research Laboratory, Navrangapura, Ahmedabad - 380009, Gujarat, India}

\author[0000-0002-2806-5821]{S. Davis}
\affil{Department of Physics, Florida State University, 77 Chieftan Way, Tallahassee, FL 32306, USA}

\author[0000-0002-0805-1908]{T. R. Diamond}
\affil{Laboratory of Observational Cosmology, Code 665, NASA Goddard Space Flight Center, Greenbelt, MD 20771, USA}

\author[0000-0001-5247-1486]{G. Folatelli} 
\affil{Facultad de Ciencias Astron\'{o}micas y Geof\'{i}sicas, Universidad Nacional de La Plata, Instituto de Astrof\'{i}sica de La Plata (IALP), CONICET, Paseo del Bosque S/N, B1900FWA La Plata, Argentina}

\author[0000-0003-3431-9135]{Wendy L. Freedman}
\affil{Department of Astronomy and Astrophysics, University of Chicago, 5640 Ellis Ave, Chicago, IL 60637, USA}

\author[0000-0003-3459-2270]{F. F\"{o}rster}
\affil{Millennium Institute of Astrophysics, Casilla 36-D,7591245, Santiago, Chile}
\affil{Departamento de Astronom\'{i}a, Universidad de Chile, Casilla 36-D, Santiago, Chile}

\author[0000-0002-1296-6887]{L. Galbany}
\affil{PITT PACC, Department of Physics and Astronomy, University of Pittsburgh, Pittsburgh, PA 15260, USA}

\author[0000-0002-8526-3963]{C. Gall}
\affil{Dark Cosmology Centre, Niels Bohr Institute, University of Copenhagen, Juliane Maries Vej 30, 2100 Copenhagen, Denmark}

\author[0000-0001-9541-0317]{S. Gonz\'alez-Gait\'an}
\affiliation{CENTRA - Centro de Astrof\'isica e Gravita\c{c}\~{a}o, Instituto Superior T\'ecnico, Av. Rovisco Pais 1, 1049-001, Lisbon, Portugal}

\author[0000-0002-4163-4996]{A. Goobar}
\affil{The Oskar Klein Centre, Department of Physics, Stockholm University, AlbaNova, 10691 Stockholm, Sweden}

\author[0000-0001-7981-8320]{M. Hamuy}
\affil{Departamento de Astronom\'{i}a, Universidad de Chile, Casilla 36-D, Santiago, Chile}

\author{S. Holmbo}
\affil{Department of Physics and Astronomy, Aarhus University, Ny Munkegade, DK-8000 Aarhus C, Denmark}

\author[0000-0002-5619-4938]{M. M. Kasliwal}
\affil{4Division of Physics, Mathematics and Astronomy, California Institute of Technology, Pasadena, California 91125, USA}

\author[0000-0002-6650-694X]{K. Krisciunas}
\affil{George P. and Cynthia Woods Mitchell Institute for Fundamental Physics and Astronomy, Department of Physics and Astronomy, Texas A\&M University, College Station, TX, 77843, USA}

\author[0000-0001-8367-7591]{S. Kumar}
\affil{Department of Physics, Florida State University, 77 Chieftan Way, Tallahassee, FL 32306, USA}

\author[0000-0003-1731-0497]{C. Lidman}
\affil{The Research School of Astronomy and Astrophysics, Australian National University, ACT 2601, Australia}

\author[0000-0002-3900-1452]{J. Lu}
\affil{Department of Physics, Florida State University, 77 Chieftan Way, Tallahassee, FL 32306, USA}

\author[0000-0002-3389-0586]{P. E. Nugent}
\affil{Lawrence Berkeley National Laboratory, Department of Physics, 1 Cyclotron Road, Berkeley, CA 94720, USA}
\affil{Astronomy Department, University of California at Berkeley, Berkeley, CA 94720, USA}

\author[0000-0002-4436-4661]{S. Perlmutter}
\affil{Lawrence Berkeley National Laboratory, Department of Physics, 1 Cyclotron Road, Berkeley, CA 94720, USA}
\affil{Astronomy Department, University of California at Berkeley, Berkeley, CA 94720, USA}

\author[0000-0003-0554-7083]{S. E. Persson}
\affil{Observatories of the Carnegie Institution for Science, 813 Santa Barbara St, Pasadena, CA 91101, USA}

\author[0000-0001-6806-0673]{A. L. Piro}
\affil{Observatories of the Carnegie Institution for Science, 813 Santa Barbara St, Pasadena, CA 91101, USA}

\author{D. Rabinowitz}
\affil{Physics Department, Yale University, 217 Prospect Street, New Haven, CT 06511, USA}

\author{M. Roth}
\affil{Carnegie Observatories, Las Campanas Observatory, Colina El Pino, Casilla 601, Chile}
\affil{GMTO Corporation, Avenida Presidente Riesco 5335, Suite 501, Las Condes, Santiago, Chile}

\author[0000-0003-4501-8100]{S. D. Ryder}
\affil{Department of Physics \& Astronomy, Macquarie University, NSW 2109, Australia}

\author[0000-0001-6589-1287]{B. P. Schmidt}
\affil{Research School of Astronomy and Astrophysics, The Australian National University, Weston, ACT 2611, Australia}

\author[0000-0002-9301-5302]{M. Shahbandeh}
\affil{Department of Physics, Florida State University, 77 Chieftan Way, Tallahassee, FL 32306, USA}

\author[0000-0002-8102-181X]{N. B. Suntzeff}
\affil{George P. and Cynthia Woods Mitchell Institute for Fundamental Physics and Astronomy, Department of Physics and Astronomy, Texas A\&M University, College Station, TX, 77843, USA}

\author[0000-0002-2387-6801]{F. Taddia}
\affil{The Oskar Klein Centre, Department of Astronomy, Stockholm University, AlbaNova, 10691 Stockholm, Sweden}

\author[0000-0002-9413-4186]{S. Uddin}
\affil{Observatories of the Carnegie Institution for Science, 813 Santa Barbara St, Pasadena, CA 91101, USA}

\author{L. Wang}
\affil{Key Laboratory of Optical Astronomy, National Astronomical Observatories, Chinese Academy of Sciences, Beijing 100012, China}
\affil{Chinese Academy of Sciences South America Center for Astronomy, China-Chile Joint Center for Astronomy, Camino El Observatorio 1515, Las Condes, Santiago, Chile}

\correspondingauthor{E. Y. Hsiao}
\email{ehsiao@fsu.edu}

\begin{abstract}
Shifting the focus of Type Ia supernova (\snia) cosmology to the
near-infrared (NIR) is a promising way to significantly reduce the
systematic errors, as the strategy minimizes our reliance on the
empirical width-luminosity relation and uncertain dust laws.
Observations in the NIR are also crucial for our understanding of the
origins and evolution of these events, further improving their
cosmological utility.  Any future experiments in the rest-frame NIR
will require knowledge of the \snia\ NIR spectroscopic diversity,
which is currently based on a small sample of observed spectra.  Along
with the accompanying paper, Phillips et al. (2018), we introduce the
Carnegie Supernova Project-II (CSP-II), to follow up nearby \sneia\ in
both the optical and the NIR.  In particular, this paper focuses on
the CSP-II NIR spectroscopy program, describing the survey strategy,
instrumental setups, data reduction, sample characteristics, and
future analyses on the data set.  In collaboration with the
Harvard-Smithsonian Center for Astrophysics (CfA) Supernova Group, we
obtained \nirspectot\ NIR spectra of \snetot\ \sneia.  Within this
sample, \nirspecinp\ NIR spectra of \sneinp\ \sneia\ have
corresponding CSP-II follow-up light curves.  Such a sample will allow
detailed studies of the NIR spectroscopic properties of \sneia,
providing a different perspective on the properties of the unburned
material, radioactive and stable nickel produced, progenitor magnetic
fields, and searches for possible signatures of companion stars.
\end{abstract}


%%%%%%%%%%%%%%%%%%
%% Introduction %%
%%%%%%%%%%%%%%%%%%

\section{Introduction}
\label{s:introduction}

%dark energy and SNe Ia

The surprising discovery of the accelerated expansion of the Universe
was based on observations of Type Ia supernovae
\citep[\sneia;][]{1998AJ....116.1009R, 1999ApJ...517..565P}.
Understanding the underlying cause of this ``dark energy'' ranks as
one of the critical tasks of contemporary physics.  \sneia\ remain the
most direct probes and when combined with complementary techniques,
such as cosmic microwave background and baryon acoustic oscillations,
provide crucial limits on the dark energy equation of state parameter
\citep[e.g.,][]{2014A&A...568A..22B, 2018ApJ...859..101S}.

%systematic errors and NIR

\snia\ cosmology has however reached an impasse.  It is currently
limited by systematic errors \citep[e.g.,][]{2011ApJS..192....1C,
  2012ApJ...746...85S} and increasing the sample size does not improve
the situation.  Despite years of significant efforts, making progress
in reducing these errors has proven difficult.  While shifting
observations to the near-infrared (NIR) is technically more
challenging (e.g., more affected by telluric absorption and fewer high
precision photometric standards), it offers a promising way forward in
two separate respects: 1) by effectively circumventing the empirical
relations that \snia\ cosmology relies upon, such as, dust laws and
luminosity-light-curve-shape relations, and 2) by systematically
exploring the NIR window to understand the physics of these
explosions.

%avoiding the unknowns

While \sneia\ are ``standardizable'' candles in the optical following
the width-luminosity relation, in the sense that fainter \sneia\ have
faster declining light curves \citep{1993ApJ...413L.105P}, they are
close to being standard candles in the NIR
\citep[e.g.,][]{2004ApJ...602L..81K, 2010AJ....139..120F,
  2012PASP..124..114K}.  Fainter \sneia\ also have lower temperatures
radiating a higher percentage of their energy at redder wavelengths.
This effect serendipitously creates a regulating mechanism, which
produces near constant peak magnitudes in the NIR
\citep{2006ApJ...649..939K}.  Shifting observations to the rest-frame
NIR reduces our reliance on an empirical width-luminosity relation.
Furthermore, color corrections due to dust and any systematic errors
associated with these are much smaller in the NIR compared to the
optical, avoiding the reliance on uncertain dust extinction laws.

%uncertain origins

There is a consensus that \sneia\ are thermonuclear disruptions of
carbon-oxygen white dwarfs.  However, beyond that, the origins of
these explosions are uncertain.  The companion star can be a
non-degenerate star: a main sequence, helium or red giant star, in a
single-degenerate system \citep{1973ApJ...186.1007W,
  1982ApJ...253..798N} or another white dwarf in a double-generate
system \citep{1984ApJS...54..335I, 1984ApJ...277..355W}.  Independent
of the progenitor systems, the triggering mechanism is also under
debate.  For example, when the mass of a carbon-oxygen white dwarf
approaches the Chandrasekhar mass, the explosion can be triggered near
the center by compressional heating.  For this mechanism, a transition
of the nuclear flame front from deflagration to detonation
\citep[``delayed detonation'' or DDT;][]{1991A&A...245..114K} appears
to match observations well \citep[e.g.,][]{1995ApJ...443...89H,
  1998ApJ...496..908W}.  Scenarios that trigger explosions in
sub-Chandrasekhar-mass white dwarfs have also been proposed.  For
example, in a ``helium detonation'' scenario, the surface He layer of
a sub-Chandrasekhar-mass white dwarf is detonated, which in turn
drives a shock wave that subsequently detonates near the center of the
white dwarf \citep[e.g.,][]{2010A&A...514A..53F}.  It is currently
unclear whether the population of \sneia\ used for cosmological
studies is composed of explosions of a single triggering mechanism
\citep{2017ApJ...846...58H} or multiple \citep{2017MNRAS.470..157B}.
Because of these uncertainties, incorporating or reducing associated
systematics is not straightforward.

%NIR

The NIR carries independent information in both light curves and
spectra \citep[e.g.,][]{2011ApJ...731..120M, 2013ApJ...766...72H}.  In
particular, it is easier to distinguish varying distributions of
intermediate-mass elements near maximum and of iron-group elements
past maximum in the NIR compared to the optical.  The NIR lines have
moderate optical depth \citep{1998ApJ...496..908W,
  2002ApJ...568..791H}, in contrast to, for example, the often
saturated Ca and Si lines in the optical
\citep[e.g.,][]{2008MNRAS.389.1087H}.  These NIR lines then offer many
clues to the physics of these explosions.  The approach of the
Carnegie Supernova Project (CSP-I) was to emphasize the NIR, and the
second phase (CSP-II) was no different.

%CSP-I

The CSP-I was an NSF-funded project to obtain optical and NIR light
curves of SNe Ia in a well-defined and understood photometric system
\citep{2006PASP..118....2H}.  It ran from 2004 to 2009, and followed
more than 123 nearby \sneia\ \citep{2010AJ....139..519C,
  2011AJ....142..156S, 2017AJ....154..211K}, as well as SNe of other
types \citep[e.g.,][]{2018A&A...609A.134S}, with rapid cadence and
high photometric precision.  With this data set, CSP-I established a
low-redshift anchor on the Hubble diagram and provided physical
insights into these explosions.  However, a sizable fraction of the
\sneia\ obtained in CSP-I was not in the smooth Hubble flow, and
therefore were susceptible to large peculiar velocity errors that are
comparable to the intrinsic dispersion we are investigating.  The
solution is to observe \sneia\ in the Hubble flow, as
\citet{2012MNRAS.425.1007B} did for a dozen \sneia\ at $0.03<z<0.09$.
In the $H$ band, they found a very encouraging result of a $0.085$ mag
scatter in the peak magnitude or $4\%$ uncertainty in distance, even
with rudimentary K-corrections applied.  More recently,
\citet{2018A&A...615A..45S} also confirmed the utility of distant
\sneia\ in the NIR, using 16 \sneia\ with single-epoch NIR photometric
observations out to a redshift of $z=0.183$.  Their K-corrections were
derived from previously published NIR spectra of 10 \sneia.

%CSP-II

CSP-II was a four-year NSF-supported follow-up program that ran from
2011 to 2015.  The main goals were to obtain optical and NIR light
curves of a ``Cosmology'' sample of \sneia\ in the smooth Hubble flow,
and to obtain optical and NIR spectra, as well as accompanying optical
and NIR light curves of a ``Physics'' sample of nearby \sneia\ to
improve the NIR K-corrections and to improve our understanding of
these events.  Several instrumental improvements were made between
CSP-I and CSP-II, which helped in these goals (see Phillips et
al. 2018 for details).  The optical imager on the 1-m Swope telescope
was upgraded.  The RetroCam NIR imager was moved from the 1-m Swope to
the 2.5-m du Pont telescope allowing observations of \sneia\ located
out in the Hubble flow.  Two instruments, both mounted on the 6.5-m
Magellan Baade telescope, were newly commissioned at the start of
CSP-II: the NIR spectrometer, the Folded-port InfraRed Echellette
\citep[FIRE;][]{2013PASP..125..270S}, and the NIR imager, FourStar
\citep{2013PASP..125..654P}.  During CSP-I, observations with the
Magellan telescopes were used to construct a rest-frame $I$-band
Hubble diagram of high-redshift \sneia\ \citep{2009ApJ...704.1036F}.
In CSP-II, the Magellan telescope observations were entirely dedicated
to the low-redshift survey, mainly to obtain NIR spectra with FIRE.
Observations ran from October to May, coinciding with the Chilean
summer, when photometric conditions are present on $>70\%$ of the
nights at the Las Campanas Observatory (LCO).

%paper organization

In this paper, we introduce the NIR spectroscopy program and the
``Physics'' sample of CSP-II; while in an accompanying paper, Phillips
et al. (2018), a general overview of the CSP-II with emphasis on the
imaging observations, is presented.  In Section~\ref{s:observations},
the observations are described, including the instruments used and the
reduction methods applied.  In Section~\ref{s:sample}, we describe our
sample.  In Sections~\ref{s:kcorr} and \ref{s:physics}, we describe
the use of this sample to improve K-corrections and our understanding
of the physics of these explosions, respectively.  A summary is then
presented in Section~\ref{s:conclusion}.

%%%%%%%%%%%%%%%%%%
%% Observations %%
%%%%%%%%%%%%%%%%%%

\section{Observations and data reductions}
\label{s:observations}

In this section, we describe the instrumental settings and data
reduction for the main instruments we used to obtain NIR spectra for
this program.  A wide range of instruments allows the follow up of
supernovae at a range of magnitudes.  The low-resolution modes of
FIRE, GNIRS, and FLAMINGOS-2 are capable of obtaining spectra with
S/N$\gtrsim20$ of objects brighter than $H\sim19-20$ mag.  The prism
mode of SpeX is capable of reaching the same S/N with objects brighter
than $H\sim 17-18$ mag.

\subsection{FIRE}
\label{ss:fire}

%intro

As seen in Table~\ref{t:nirspec}, FIRE at the LCO 6.5-m Magellan Baade
Telescope was the main workhorse for the CSP-II NIR spectroscopy
program.  It was newly commissioned and released for general use in
the 2010B semester.  Three nights of test runs were conducted in 2011A
to determine the optimal observational set up for supernovae.  Several
\sneia\ were observed during the test runs, including the
SN~2002cx-like SN~2011ce, although these test observations do not have
accompanying CSP-II light curves.  The CSP-II began in the 2011B
semester.

%observational set up

The majority of FIRE spectra were obtained in the high-throughput
prism mode with a 0\farcs6 slit.  This configuration yields continuous
wavelength coverage from 0.8 to 2.5 \um\ with resolutions of
$R\sim500$, $450$, and $300$ in the $JHK$ bands, respectively.  Only a
handful of nearby \sneia\ (within 20 Mpc) were observed in the
high-resolution echellette mode, which yields a resolution of
$R=6000$.  When acquiring a target, the slit was oriented along the
parallactic angle to minimize the effect of differential refraction
\citep{1982PASP...94..715F}.  For each science observation, an A0V
star close to the science observation in time, angular distance and
air mass, was observed as the telluric and flux standard.  For each
science or telluric observation, several frames ($>4$) were obtained
using the conventional ABBA ``nod-along-the-slit'' technique.  The
``sampling-up-the-ramp'' readout mode was chosen for the science
exposures to reduce the readout noise through the sampling of multiple
non-destructive reads.  The per-frame exposure time was typically
$\sim2$ minutes or shorter, depending on the brightness of the
supernova.  These exposure times were chosen such that an adequate
signal was obtained in each frame without saturating the detector with
airglow emissions in the $JHK$ bands and thermal background, mostly
affecting the red side of $K$ band.  For each science and telluric
observation, a spectrum was taken of Ne and Ar arc lamps for
wavelength calibration.  The ``low gain'' mode of 3.8 e$-$/DN for the
detector gain setting was always chosen to avoid saturation of the A0V
telluric standards which were typically between 10 to 12 mag.

%FIRE reductions

The data were reduced using the IDL pipeline \texttt{firehose}
\citep{2013PASP..125..270S}, specifically designed for the reduction
of FIRE data.  The pipeline performed steps of flat fielding,
wavelength calibration, sky subtraction, spectral tracing, and
extraction.  Wavelength calibration was done using $\sim40$ Ne/Ar
lines, which are evenly spaced out in pixel space.  The wavelength
solutions typically yielded $2-6$ \AA\ rms dispersion from the blue
end to the red end, and a $4$ \AA\ rms dispersion overall.  For the
removal of sky lines and background, the background flux was modeled
using off-source pixels as described by \citet{2003PASP..115..688K}
and subtracted from each frame, instead of the conventional A-B pair
subtractions.  This step removed the host galaxy background in the
cases of moderate host contamination.  No further steps were taken to
subtract host galaxy light.  The spectral extraction was then
performed using the optimal extraction technique
\citep{1986PASP...98..609H}, a weighting scheme that maximizes S/N
ratio while preserving spectrophotometric accuracy.  We took advantage
of the multiple spectra to perform sigma clipping to reject spurious
pixels, and also to produce an error spectrum by computing the
standard deviation at each pixel.  Individual 1D spectra were then
median combined.  Corrections for telluric absorptions were performed
using the IDL tool \texttt{xtellcor} developed by
\citet{2003PASP..115..389V}.  To construct a telluric correction
spectrum free of stellar absorption features, a model spectrum of Vega
was used to match and remove the hydrogen lines of the Paschen and
Brackett series from the A0V telluric standard.  The resulting
telluric correction spectrum was also used for flux calibration, given
the magnitudes of the A0V star from the Simbad database.

%spectrophotometric accuracy

We used the FIRE spectra of~SN 2012fr and its $YJH$ light curves from
\citet{2018ApJ...859...24C} to assess the spectrophotometric accuracy
of our flux calibration.  With 13 epochs of simultaneous NIR spectra
and light curves, we obtained a median of the differences between the
colors from spectra and photometry of 0.08 and 0.03 mag for each of
$Y-J$ and $J-H$ colors, respectively.  This is comparable to the
precision obtained in optical spectra
\citep[e.g.,][]{2012MNRAS.425.1789S}.  The spectrophotometric errors
in the NIR are expected to be larger since the A0V standard stars are
used as both telluric and flux standards.

%resolution

In the high-throughput mode of FIRE, since the disperser includes
prisms, the spacing in wavelength is not constant; rather, the
logarithm of wavelength yields approximately constant spacing.  As a
consequence, the spectral resolution drops from blue to red.  Note
however that the detector pixels are efficiently used, since the
supernova features are sampled roughly constantly in velocity space
throughout the entire 0.8 to 2.5 \um\ wavelength coverage.  Because of
the non-linear wavelength spacing, we chose not to record the
wavelength vector of a prism with only the first wavelength element
and a wavelength spacing given in the header.  Instead, the spectra
were saved with wavelength, flux, and flux error at each of the 2048
detector pixels.

%Quick reduction

A quick reduction pipeline was also developed for the FIRE
high-throughput mode and installed at the Magellan Baade Telescope.
The quick reduction pipeline is a wrapper to the \texttt{firehose}
pipeline, and uses archival flats, arcs and telluric spectra to
automate the entire process.  Since \texttt{firehose} does not require
AB pairs for background subtraction, a quick spectrum can be produced
upon the completion of a single observation.  Processing of a single
frame takes $\sim1.5$ minutes.  This is shorter than the typical per
frame exposure time of $\sim2$ minutes (without overhead).  We could
therefore process each frame on the fly.  The 1D spectra from ABBA
frames were then stacked.  The quick reduction spectra typically have
poor telluric correction, as there is currently no sophisticated
algorithm to select or adjust an archival telluric spectrum.

%Advantages

The quick reduction pipeline serves several important functions.  It
offers the observer a check of the S/N ratio after each successive
frame and eliminates the guesswork in the number of frames required to
reach the desired S/N ratio.  This has been shown to be a substantial
time saving measure and increased the number of targets observed per
night by two- to three-fold.  The quick reduction also offers the
observer a check of the identity of the target under the slit within
the first two frames of observation.  In the case of a supernova, the
features are identifiable and easily distinguished from the spectra of
foreground field stars.  Thus, the quick reduction also allows the
efficient classification of newly discovered supernovae.  Due to the
rarity of early-time NIR spectra of supernovae and the valuable
information they contain on the outer ejecta, we have made a concerted
effort to target newly discovered and often unclassified supernovae
with FIRE.  With the quick reduction pipeline, FIRE becomes an
efficient classification machine.  Typically, a spectrum is produced
from the pipeline within five minutes of acquiring the target.  The
S/N ratio for a single-frame NIR spectrum is usually not adequate for
science, but is often adequate for classification.  Based on this, we
can then decide to stay on the target or move on if the supernova does
not meet our criteria for follow up.  In four years, 44 supernovae
were classified using FIRE by CSP-II.  The \sneia\ classified by FIRE
are listed in Table~\ref{t:nirspec_name}.  All classifications were
immediately reported to either or both the Astronomer's Telegram
(ATel) and the Central Bureau Electronic Telegrams (CBET).

\subsection{GNIRS}
\label{ss:gnirs}

%GNIRS observations

Spectra observed with the Gemini Near InfraRed Spectrograph
\citep[GNIRS;][]{2006SPIE.6269E..4CE} on the 8.2-m Gemini North
telescope were taken in the cross-dispersed mode, in combination with
the short-wavelength camera, a 32 lines per mm grating, and
0$\farcs$675 slit.  This configuration allowed for a wide wavelength
coverage from 0.8 to 2.5 \um, divided over six orders with a
resolution of $R\sim1000$.  The observing setup was similar to that
described for FIRE observations.  Because of the higher resolution for
GNIRS, longer per-frame exposure times were used when necessary,
between 240 and 300 seconds.  The slit was positioned at the
parallactic angle at the beginning of each observation.  An A0V star
was also observed before or after each set of science observations for
telluric and flux calibration.

%GNIRS reduction

The GNIRS data were calibrated and reduced using the \texttt{XDGNIRS}
pipeline, specifically developed for the reduction of GNIRS
cross-dispersed data.  The pipeline is partially based on the
\texttt{REDCAN} pipeline for reduction of mid-IR imaging and
spectroscopy from CANARICAM on the Gran Telescopio Canarias
\citep{2013A&A...553A..35G}.  The steps began with pattern noise
cleaning, non-linearity correction, locating the spectral orders and
flat-fielding.  Sky subtractions were performed for each AB pair
closest in time, then the 2D spectra were stacked.  Spatial distortion
corrections and wavelength calibrations were applied before the 1D
spectrum was extracted.  Corrections for telluric absorption, and
simultaneous flux calibration, were done with the \texttt{XTELLCOR}
software package and the A0V telluric standard observations, using the
prescription of \citet{2003PASP..115..389V}.  As a final step, the six
orders were joined to form a single continuous spectrum.

\subsection{FLAMINGOS-2}
\label{ss:flamingos}

The FLAMINGOS-2 \citep{2008SPIE.7014E..0VE} data were taken with the
Gemini South 8.2-m telescope in long-slit mode with the $JH$ grism and
filter in place, in most cases using a 0\farcs72 slit width.  This
setup yielded a wavelength range of $1.0-1.8$ \um\ and $R\sim1000$.
Again, the data were taken at the parallactic angle with a standard
ABBA pattern, and with typical per-frame exposure times of $100-300$
s.  These long-slit data were reduced in a standard way using the
\texttt{F2 PyRAF} package provided by Gemini Observatory, with image
detrending, sky subtraction of the AB pairs, spectral extraction,
wavelength calibration and spectral combination.  Telluric corrections
and flux calibrations were again determined using the
\texttt{XTELLCOR} package.

\subsection{SpeX}
\label{ss:spex}

The SpeX \citep{2003PASP..115..362R} data obtained with the 3.0-m NASA
Infrared Telescope Facility (IRTF) were generally observed in the
so-called ``SXD'' mode, where the spectrum is cross-dispersed to
obtain wavelength coverage from $\sim0.8-2.4$ \um\ in a single
exposure, spread over six orders.  This setup yielded $R\sim1200$ with
the 0\farcs5 slit, the one most often used.  All observations were
taken with the classic ABBA technique, with typical per-frame
exposures of 150 s.  Further, all observations were taken with the
slit oriented along the parallactic angle.  As with the NIR data from
other spectrographs presented in this work, a A0V star was observed
before or after the science observations for flux and telluric
calibration.

All SpeX data were reduced using the publicly available
\texttt{Spextool} software package \citep{2004PASP..116..362C}.  This
reduction proceeded in a standard way, with image detrending, order
identification and sky subtraction using the AB pairs closest in time.
Corrections for telluric absorption utilized the \texttt{XTELLCOR}
software and A0V star observations \citep{2003PASP..115..389V}.  After
extraction and telluric correction, the 1D spectra from the six orders
were rescaled and combined into a single spectrum.

%%%%%%%%%%%%
%% Sample %%
%%%%%%%%%%%%

\section{Sample Characteristics}
\label{s:sample}

%physics vs cosmology

We obtained NIR spectra for \sneia\ in the CSP-II Physics sample
to further our understanding of the origins of these explosions and to
improve NIR K-corrections.  The main difference between the
Cosmology and Physics samples is the host galaxy redshifts.
We selected \sneia\ in the Hubble flow for the Cosmology sample,
while \sneia\ in the Physics sample were required to be nearby or
bright enough for high signal-to-noise (S/N), NIR spectroscopic
follow up, at least until approximately one month past maximum.  For a
normal \snia\ this effectively placed an upper redshift limit at the
edge of the Hubble flow, $z\sim 0.03-0.04$.  Thus, the Cosmology
and Physics samples are mostly independent, although there are
\sneincp\ objects which belong to both samples.  In total, there are
\sneinc\ and \sneinp\ objects in the Cosmology and Physics
samples, respectively.  Details of the NIR spectroscopy sample are
tabulated in Table~\ref{t:nirspec_name}, and the details of the
Cosmology and Physics samples can be found in Phillips et
al. (2018).

%other differences

There are other differences between the samples.  For \sneia\ in the
Physics sample, we obtained the $BV$ and $ugri$ light curves with the
Swope telescope and $YJH$ with du Pont and Magellan whenever possible;
whereas for those in the Cosmology sample, observations for $u$, $g$
and sometimes $H$ bands were excluded, since these objects required
considerably longer exposure times.  \sneia\ in the Cosmology sample
were drawn almost entirely from ``blind'' transient searches to
minimize any bias toward more massive galaxies.  These were wide-field
and untargeted searches, such as the Palomar Transient Factory
\citep[PTF;][]{2009PASP..121.1395L}, the intermediate Palomar
Transient Factory (iPTF) and the La Silla Quest Low-Redshift Supernova
Survey \citep{2013PASP..125..683B}, as opposed to targeted searches
where a sample of pre-selected galaxies was monitored.  \sneia\ from
the Physics sample were drawn from both targeted (31 \sneia) and
untargeted (59 \sneia) searches.  For both samples, we generally
required that the follow-up observations began before maximum light in
the optical.  The first estimation of the phase of a \snia\ was
usually determined from the classification spectrum using tools like
\texttt{SNID} \citep{2007ApJ...666.1024B}, \texttt{superfit}
\citep{2005ApJ...634.1190H} or \texttt{GELATO}
\citep{2008A&A...488..383H}, but can sometimes be determined earlier
with rising or falling Swope light curves.

%desired characteristics

The previous largest sample of NIR spectra of \sneia\ came from the
pioneering study of \citet{2009AJ....138..727M}, which consists of 41
spectra obtained from SpeX on the NASA Infrared Telescope Facility
(IRTF).  Our goal was to improve upon this data set in several key
aspects:

\begin{enumerate}
\item \emph{Larger sample size.} \citet{2007ApJ...663.1187H}
  characterized the mean optical spectral properties of \sneia\ with
  $\sim 1000$ spectra.  The sample size of NIR spectra before CSP-II
  was two orders of magnitude smaller than that of optical spectra.
  To not only characterize the mean NIR spectral properties, but also
  to determine the variation with light-curve shape, the sample size
  of NIR spectra needed to be drastically increased.
\item \emph{Time-series observations.} Understanding the time
  evolution of NIR spectral features is important for both
  K-correction and physics studies.  The \citet{2009AJ....138..727M}
  sample is largely composed of single-epoch ``snapshots,'' and is
  valuable for diversity studies, but contains little time evolution
  information.
\item \emph{Complementary light-curve observations.} Peak absolute
  magnitudes and light-curve decline rates are key observables of
  \sneia.  However, complementary light-curve observations are often
  missing for \sneia\ in the previous sample.
\item \emph{Simultaneous optical spectra.} Optical and NIR spectra at
  the same phase probe different regions in the ejecta
  \citep{1998ApJ...496..908W, 2002ApJ...568..791H} and help confirm
  identities of the chemical elements
  \citep[e.g.,][]{2013ApJ...766...72H}.  Simultaneous optical and NIR
  spectra are also lacking in the previous sample.
\item \emph{Improved telluric regions.} Previous NIR spectra are often
  marred by the heavy telluric absorptions between the $JHK$ bands,
  mainly caused by water vapor.  These wavelength regions are crucial
  for computing accurate K-corrections (see Section~\ref{s:kcorr}).
  The key to improved telluric corrections is increased signal in the
  telluric regions.  Large telescope apertures and high-throughput
  instruments with medium resolution allow for such an improvement.
\end{enumerate}

We outline below how we achieved these improvements over the previous
sample.

%observing strategy

\begin{figure}
\centering
\includegraphics[width=0.5\textwidth,clip=true]{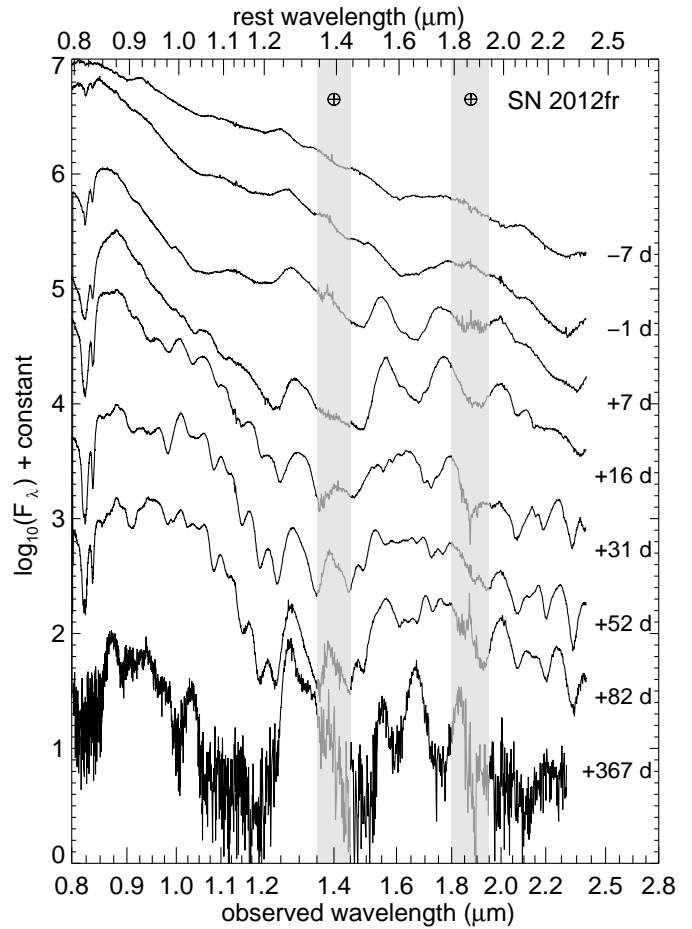}
\caption{A sampling of NIR spectroscopic time evolution of the Type Ia
  SN~2012fr from pre-maximum through to transitional and nebular
  phases.  All spectra shown were obtained with FIRE through the
  CSP-II NIR spectroscopy program.  The phase relative to $B$-band
  maximum is labeled for each spectrum.  The gray vertical bands mark
  the regions of the strongest telluric absorptions.}
\label{f:12fr}
\end{figure}

FIRE was the main workhorse of the CSP-II NIR spectroscopy program.
In the prism mode, it is capable of obtaining high S/N ratio ($\gtrsim
20$) spectra down to $H\sim19$ mag for early-phase spectra, and down
to $\sim20$ mag for the emission-line dominated, nebular-phase
spectra.  The high throughput of FIRE means that we were able to
obtain enough counts under the strong telluric absorption lines to
enable telluric corrections in most cases.  For example, we recovered
the \palpha\ P-Cygni feature, which is located in the region of heavy
telluric absorption between $H$ and $K$ bands, with 10\% precision or
better for 70\% of spectra in our Type II sample.  In
Figure~\ref{f:12fr}, we show an example NIR time-series observation of
SN~2012fr from pre-maximum to nebular phases, obtained entirely with
FIRE.  At LCO, the observing time is allotted as scheduled nights.  In
collaboration with the Harvard-Smithsonian Center for Astrophysics
(CfA) Supernova Group, we were awarded 70 Baade nights in total during
CSP-II.  That is on average 3 nights per month and allowed continuous
time-series coverage of a large sample of \sneia\ during each Chilean
summer.  The time series was then supplemented by target of
opportunity (ToO) queue mode observations from Gemini, IRTF, and Very
Large Telescope (VLT), especially during early phases.  Whenever
possible, simultaneous optical spectroscopy was obtained mainly at du
Pont, Magellan and the Nordic Optical Telescope (NOT).  The vast
majority of the \sneia\ had rapid-cadence (nightly) optical light
curve follow up with Swope before or at the beginning of spectroscopic
observations.  NIR light curve points were obtained whenever possible
with du Pont and Magellan.  For a few northern objects, such as
SNe~2011fe and 2014J, we relied entirely on northern facilities with
NIR spectroscopic capability: Gemini North, IRTF, and the Mt. Abu
1.2-m Infrared Telescope.  These objects have no complementary light
curves from CSP-II.

%stats

\begin{figure}
\centering
\includegraphics[width=0.50\textwidth,clip=true]{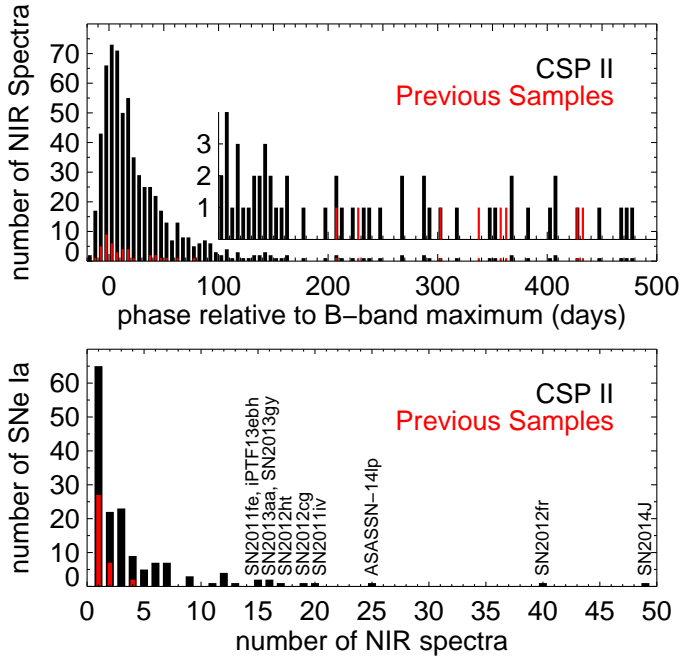}
\caption{Number of NIR spectra at each epoch relative to $B$-band
  maximum (top panel) and number of \sneia\ with/without time-series
  observations (bottom panel).  The inset of the top panel provides a
  zoomed-in view of the numbers at late times.  \sneia\ with 15 NIR
  spectra or more have their names labeled in the bottom panel.  The
  NIR spectra of SN~2011fe \citep{2013ApJ...766...72H} and iPTF13ebh
  \citep{2015A&A...578A...9H}, and a subset of the NIR spectra of
  SN~2011iv \citep{2018A&A...611A..58G}, ASASSN-14lp and SN~2014J
  \citep{2015ApJ...798...39M, 2016ApJ...822L..16S} were previously
  published.}
\label{f:snstat}
\end{figure}

\begin{figure*}
\centering
\includegraphics[width=1.0\textwidth,clip=true]{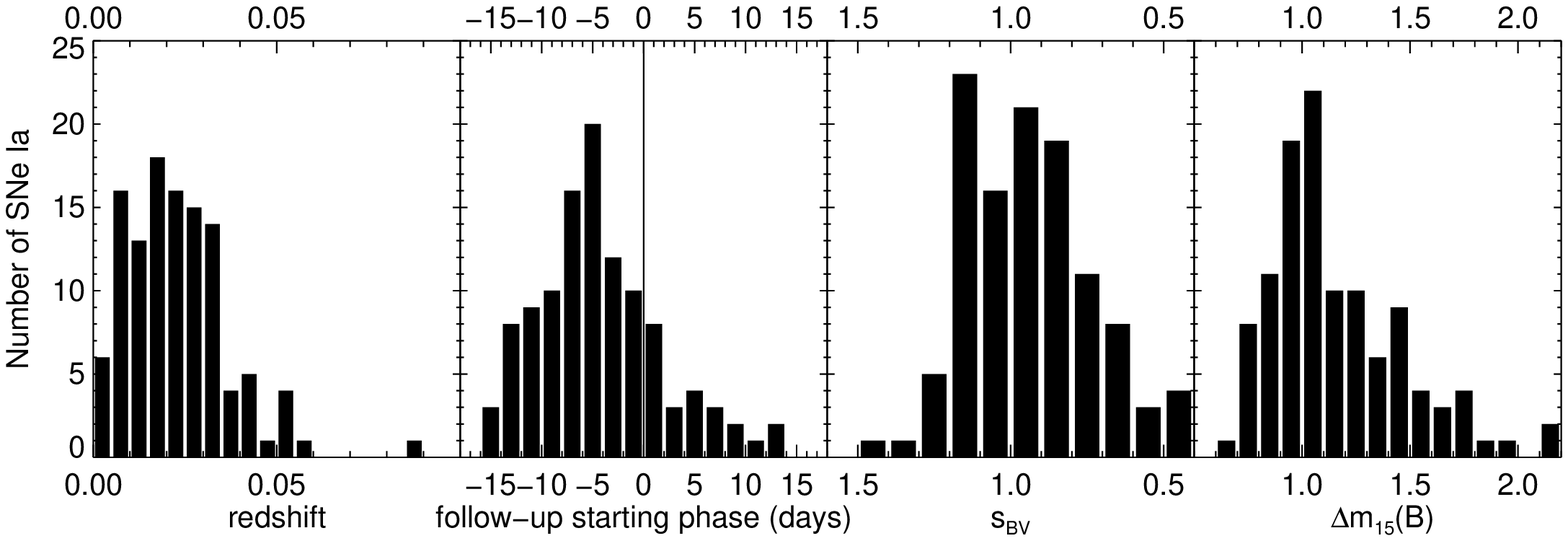}
\caption{Characteristics of the \sneia\ in the NIR spectroscopy
  program.  The distributions in redshift, light curve follow up
  starting phase, $s_{BV}$, and \dm\ are shown in panels from left to
  right.}
\label{f:snstat_sn}
\end{figure*}

The top panel of Figure~\ref{f:snstat} shows that there are many
spectra at all phases relevant for K-corrections and template building
($\lesssim 100$ days past maximum).  For a few very nearby \sneia, the
time series extends to the nebular phase, providing unique physical
diagnostics which are described in Section~\ref{s:physics}.  The
bottom panel shows that the majority of \sneia\ have two spectra or
more, providing the crucial time evolution information important for
both explosion physics and K-correction studies.
Figure~\ref{f:snstat_sn} illustrates the characteristics of the
\sneia\ observed in the NIR spectroscopy program.  Nearly all of the
\sneia\ are located within $z<0.05$.  A few more distant objects have
NIR spectra taken for classification only.  The vast majority of the
\sneia\ have optical light-curve follow up starting before maximum
light, and the sample contains \sneia\ in the full range of
light-curve shape parameters \sbv\ \citep{2014ApJ...789...32B} and
\dm\ \citep{1993ApJ...413L.105P}, determined using the updated version
of \texttt{SNooPy} \citep{2011AJ....141...19B}.  Altogether, we
obtained \nirspectot\ NIR spectra of \snetot\ \sneia.  Within this
sample, \nirspecinp\ NIR spectra of \sneinp\ \sneia\ are in the
Physics sample with CSP-II light curve follow up.

%exceptions to physics sample

The list of \sneia\ for which we obtained NIR spectra is given in
Table~\ref{t:nirspec_name}.  Note that there are several objects which
are only in the Cosmology sample.  These objects are the most distant
objects in our sample, and thus in most cases only have single-epoch
low-S/N-ratio NIR spectra.  Several objects belong to neither the
Cosmology nor Physics samples for the following reasons.  Some objects
are of the peculiar SN~2002cx-like \citep[e.g.][]{2003PASP..115..453L,
  2013ApJ...767...57F} or ``super-Chandrasekhar''
\citep[e.g.][]{2006Natur.443..308H} subtypes, which were excluded from
both samples.  Other objects do not have CSP-II follow up light curves
either because they exploded between observing campaigns (e.g.,
SN~2012cg) or they are northern objects, as mentioned previously
(e.g., SN~2011fe).

%SN of other types

Time-series NIR spectroscopy of other SN types was also obtained,
given that the existing samples are even smaller than that of
\snia\ and the diagnostics in the NIR are just as powerful.  Many NIR
spectra of various SN types obtained by CSP-II have already been
published in single-object papers: the rebrightening of SN~2009ip
\citep{2014ApJ...780...21M}, the Type IIb SN~2011hs
\citep{2014MNRAS.439.1807B}, the broad-lined Type Ic SN~2012ap
\citep{2015ApJ...799...51M}, the Type Ib SN~2012au
\citep{2013ApJ...770L..38M}, the Type II SN~2012aw
\citep{2014ApJ...787..139D}, the Type IIn SN~2012ca
\citep{2015MNRAS.447..772F}, the Type IIL SN~2013by
\citep{2015MNRAS.448.2608V}, the Type Ib/c SN~2013ge
\citep{2016ApJ...821...57D}, as well as the SN~2002cx-like SN~2012Z
\citep{2015A&A...573A...2S} and SN~2014ck \citep{2016MNRAS.459.1018T}.
Note that SN~2012ca may be a \snia\ interacting with its surrounding
circumstellar medium as noted by several studies
\citep{2014MNRAS.437L..51I, 2016MNRAS.459.2721I, 2015MNRAS.447..772F}

\begin{figure}
\centering
\includegraphics[width=0.5\textwidth,clip=true]{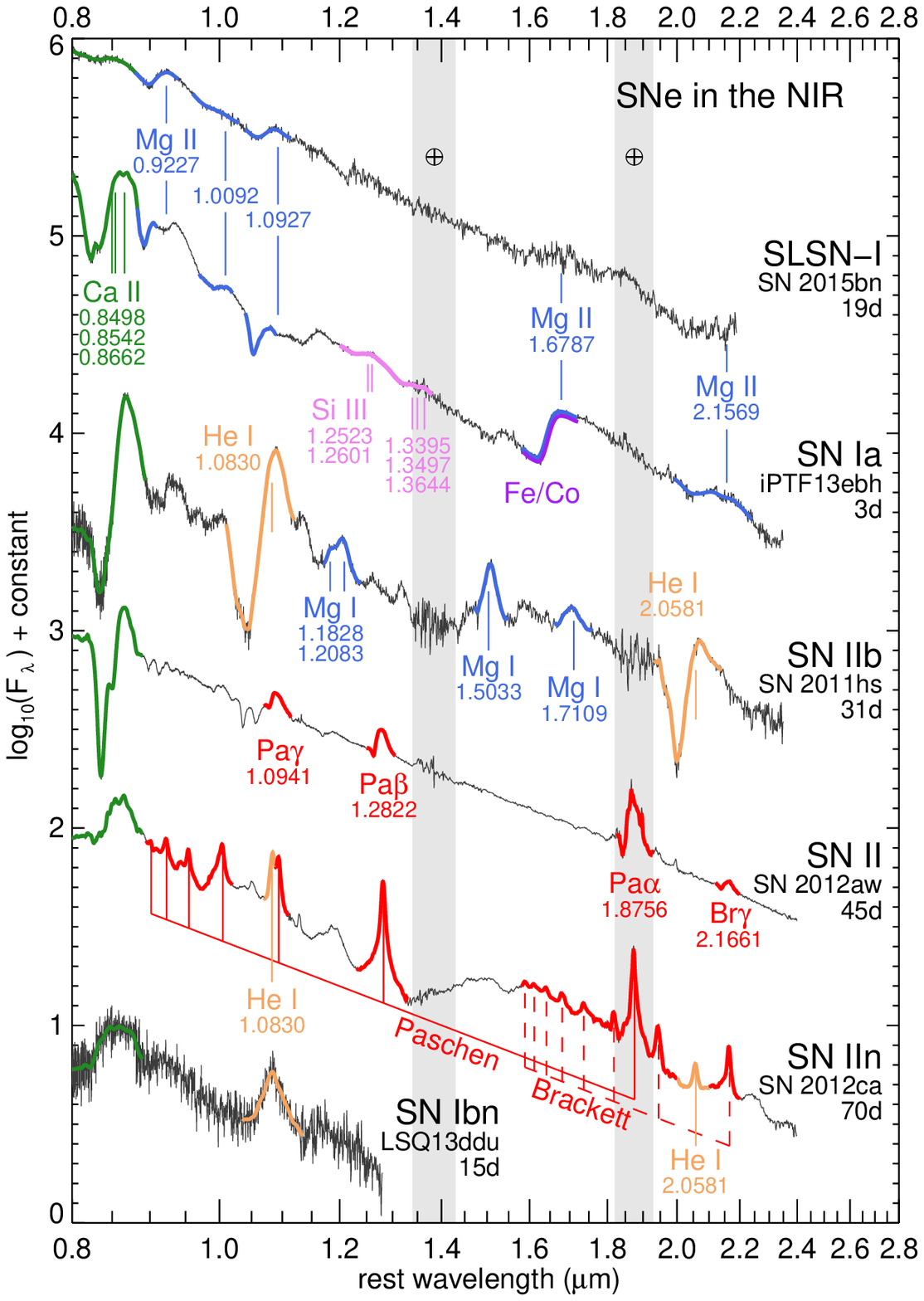}
\caption{An atlas of NIR spectra of various SN types.  The dominant
  ion species are labeled for comparison.  The laboratory wavelengths
  are labeled in microns for reference and also marked on the figure
  as vertical lines.  The supernova name and phase are labeled for
  each spectrum.  The phases are relative to maximum light, except for
  that of SN~2012aw, which is relative to explosion.  The gray
  vertical bands mark the regions of the strongest telluric
  absorptions.  All spectra shown were taken with FIRE as part of the
  CSP-II NIR spectroscopy program.  Spectra of iPTF13ebh
  \citep{2015A&A...578A...9H}, SN~2011hs \citep{2014MNRAS.439.1807B},
  SN~2012aw \citep{2014ApJ...787..139D}, and SN~2012ca
  \citep{2015MNRAS.447..772F} were published previously.  FIRE spectra
  of the SLSN-I SN~2015bn and the Type Ibn LSQ13ddu are shown for the
  first time.}
\label{f:type}
\end{figure}

An atlas of the NIR spectra of supernovae of various types is plotted
in Figure~\ref{f:type}.  The dominant ion species are labeled.  The
strong \caii\ infrared triplet is present in all types of supernovae.
Hydrogen Paschen and Brackett series are present in Type II
supernovae.  Note that \palpha\ is located in the heavy telluric
absorption between the $H$ and $K$ bands, therefore \pbeta\ is often
selected as the primary diagnostic in the NIR.  The two strong NIR
\hei\ lines are present in Type IIb, IIn, and Ibn supernovae.  One
should be cognizant of the proximity between the \hei\ \lam1.0830
\um\ line and \pgamma\ when identifying features in the wavelength
region near 1.1 \um.  Multiple \mgii\ lines are dominant in early Type
Ia and superluminous supernova (SLSN) NIR spectra.
Table~\ref{t:nirspec} shows the total number of NIR spectra taken by
CSP-II according to SN type and instrument used.  Note that these
numbers include classification spectra that do not have accompanying
light curves.  We discuss our use of FIRE for classification of SN
types further in Section~\ref{s:observations}.  For every SN type,
this data set of NIR spectra will at least double the existing
published sample; and we hope, by opening the NIR window, the data set
will further the understanding of each type of supernova.

%%%%%%%%%%%
%% Kcorr %%
%%%%%%%%%%%

\section{K-corrections}
\label{s:kcorr}

\begin{figure}
\centering
\includegraphics[width=0.5\textwidth,clip=true]{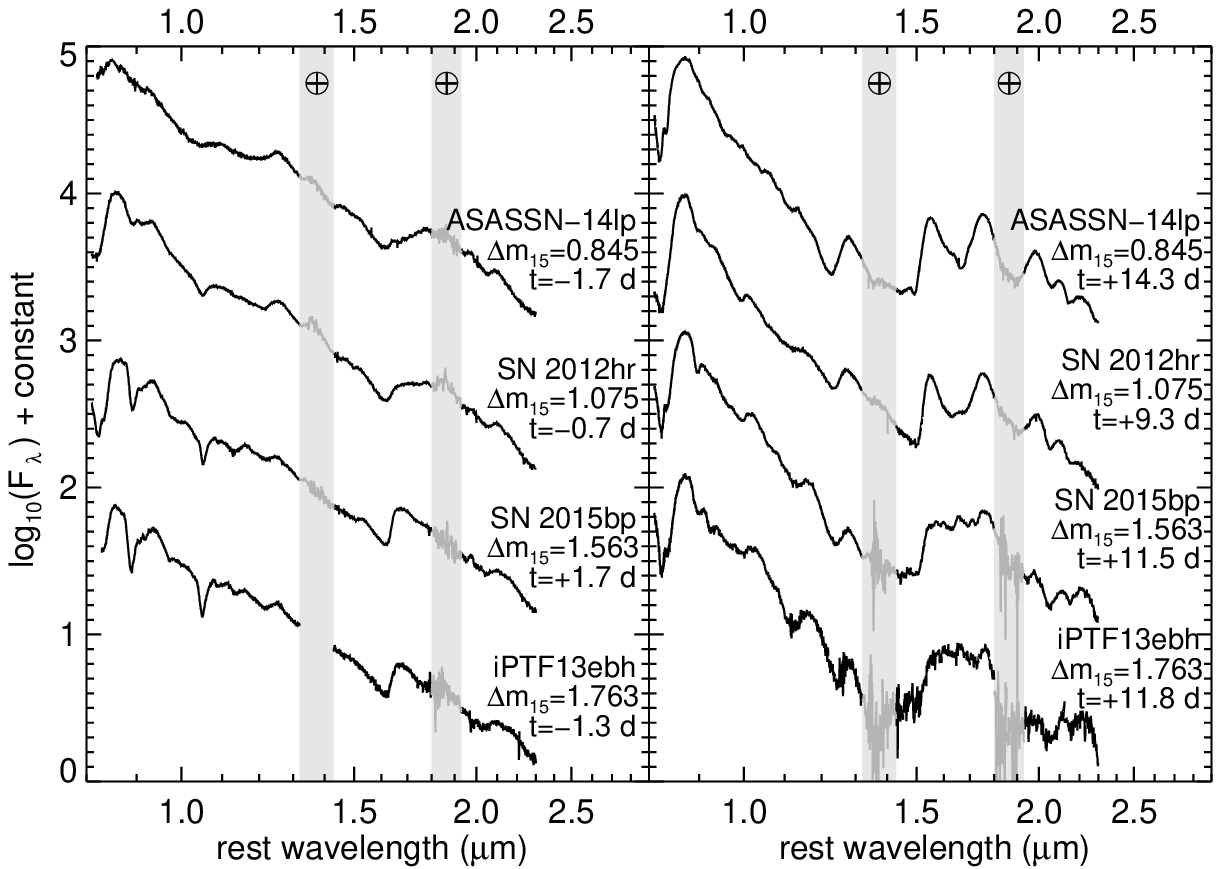}
\caption{Comparison of \snia\ NIR spectra of various decline rate
  \dm\ values.  The left and right panels show the spectra near
  maximum and $1-2$ weeks past maximum, respectively.  The SN name,
  \dm, and spectral phase are labeled for each spectrum.  The gray
  vertical bands mark the regions of the strongest telluric
  absorptions.  The spectral diversity in the NIR is strongly
  correlated with light-curve shape.}
\label{f:max12}
\end{figure}

%motivation

K-corrections account for the effect of cosmological expansion upon
measured magnitudes \citep{1968ApJ...154...21O}.  As mentioned in the
introduction, shifting \snia\ cosmology to the rest-frame NIR has the
crucial advantages of largely circumventing the empirical
width-luminosity relation and any uncertain dust reddening laws.  This
method is considered one of the most promising ways to reduce current
\snia\ cosmology systematics.  The potential of \sneia\ as standard
candles in the NIR has been demonstrated in many studies
\citep[e.g.,][]{2004ApJ...602L..81K, 2008ApJ...689..377W,
  2010AJ....139..120F, 2012MNRAS.425.1007B, 2018A&A...615A..45S, B18}.
Furthermore, all these studies utilized rudimentary K-corrections
based on a small number of spectra.  Any low-redshift experiment
testing the efficacy of \sneia\ as standard candles in the NIR and any
high-redshift experiment in the rest-frame NIR constraining the
properties of dark energy will benefit from accurate knowledge of the
diversity and time-evolution of NIR spectra.

%telluric correction

Preliminary studies of NIR K-corrections yielded the promising result
of a K-correction error from the diversity in spectral features of
0.04 mag at $z=0.08$ in the $Y$ band \citep{2014PASP..126..324B}, but
studies of the other NIR bands were limited by poor telluric
corrections.  It is only feasible to obtain high S/N ratio NIR spectra
for \sneia\ at $z\lesssim0.03$.  This redshift range does not allow
for the sampling of the entire telluric-obscured regions.
Furthermore, our CSP-II Cosmology sample for testing the efficacy of
\snia\ cosmology in the NIR has a redshift range of $0.03 \lesssim z
\lesssim 0.1$.  This means we are relying solely on the
telluric-obscured region of our NIR spectroscopic sample to K-correct
the NIR light curves of our Cosmology sample, again emphasizing the
importance of telluric corrections.  For our FIRE sample, the
high-throughput prism mode allows for the collection of enough SN
photons under the water vapor absorptions in most cases (see
Section~\ref{s:observations} for details).

%spectral variation

Another key to successful K-corrections is our ability to predict the
correct spectral energy distribution with only photometric
information.  It has been shown that even though there is substantial
spectroscopic diversity in both the optical and NIR, defining a mean
spectrum eliminates systematic errors and reduces the statistical
errors to an acceptable level \citep{2007ApJ...663.1187H,
  2009PhDT.......228H}.  There is also strong evidence that optical
spectral features vary with decline rate
\citep[e.g.,][]{1995ApJ...455L.147N}, and indications of the same
behavior in the NIR \citep{2013ApJ...766...72H}.  Figure~\ref{f:max12}
shows that \dm\ is a strong indicator of NIR spectroscopic diversity
for both near- and post-maximum phases.  For example, the strong
\mgii\ \lam1.0927\um\ absorption feature increases in strength as we
go from slow to fast-declining \sneia\ (left panel of
Figure~\ref{f:max12}).  The most prominent feature in the NIR for
\sneia, the $H$-band break near 1.5 \um, is shown quantitatively to
correlate tightly with \dm\ and $s_{BV}$ \citep[right panel of
  Figure~\ref{f:max12};][]{2013ApJ...766...72H, 2015A&A...578A...9H}.
The accuracy of K-corrections can thus be further improved by
characterizing the spectral variation with \dm\ or $s_{BV}$.  Our
large sample of NIR spectra will, for the first time, adequately fill
the parameter space of wavelength (including telluric obscured
regions), phase, and light-curve shape parameters.

%%%%%%%%%%%%%
%% Physics %%
%%%%%%%%%%%%%

\section{Physics of SN\MakeLowercase{e} I\MakeLowercase{a}}
\label{s:physics}

\begin{figure*}
\centering
%\figurenum{text}
\epsscale{1.1}
\plottwo{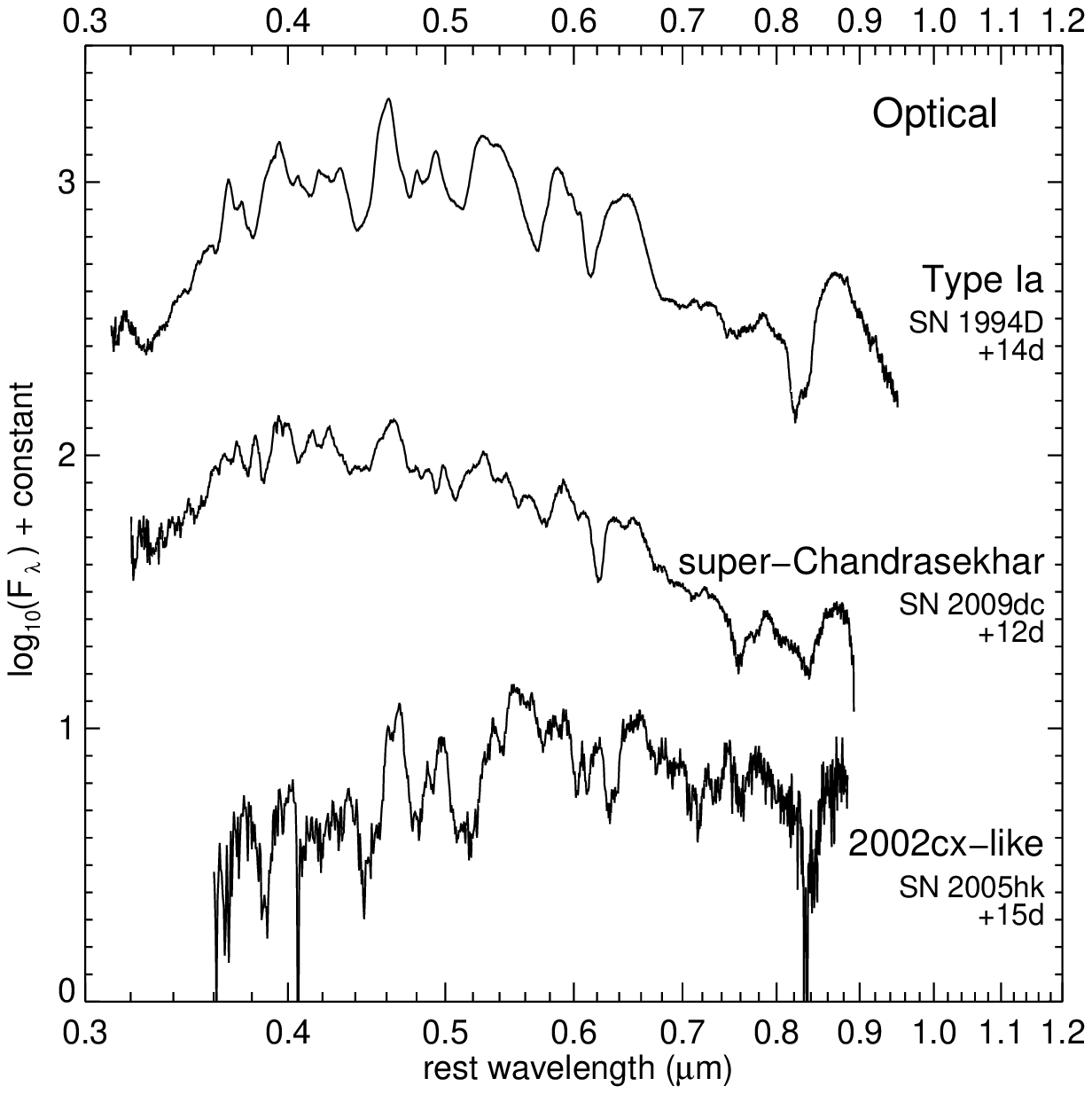}{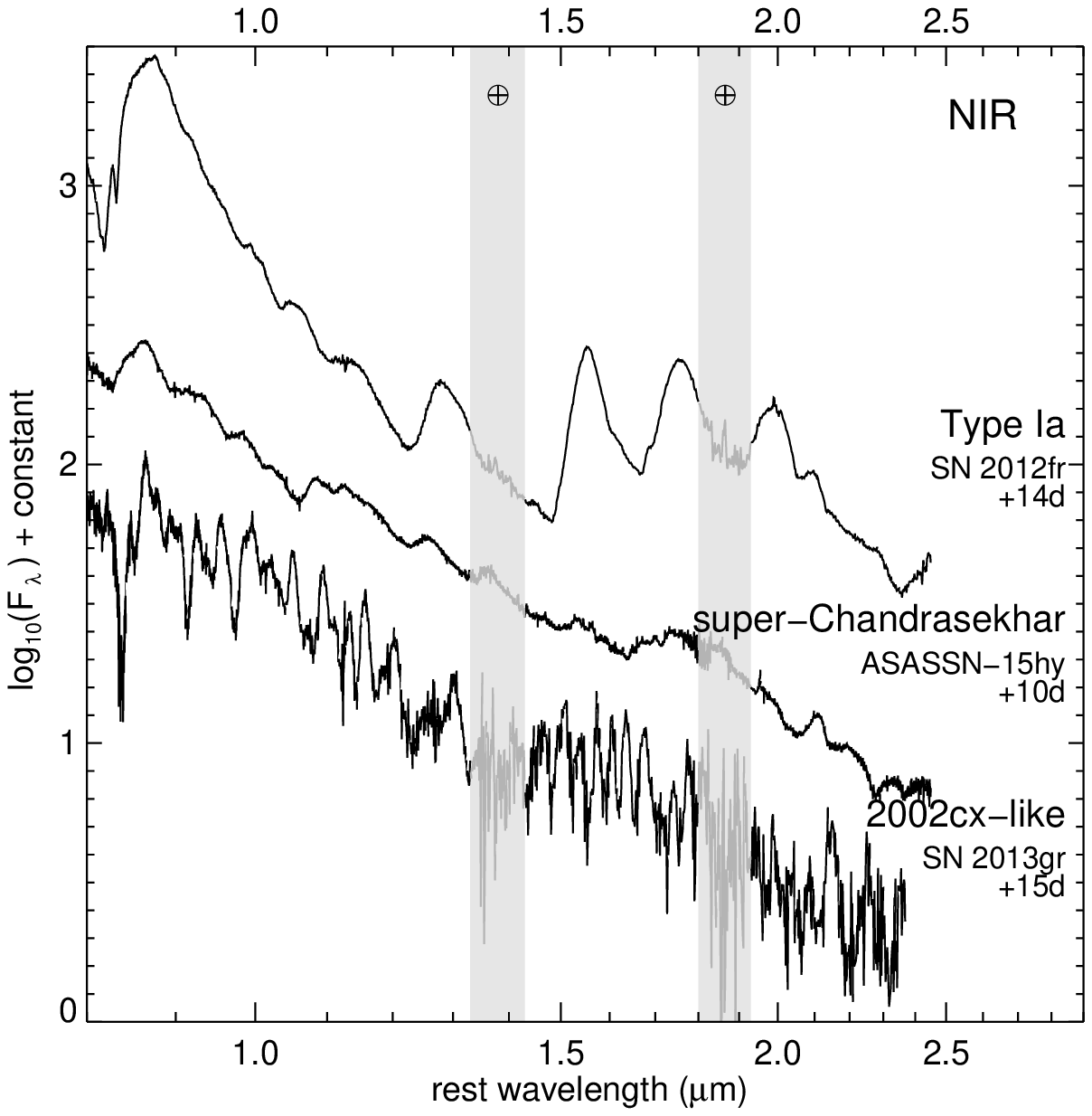}
\caption{Comparison of spectral features of normal and peculiar
  \sneia.  Left and right panels show spectra in the optical and NIR,
  respectively.  In the right panel, the gray vertical bands mark the
  regions of the strongest telluric absorptions.  In the optical, the
  spectral features between normal and peculiar \sneia\ are broadly
  similar.  Indeed, that is why they are all classified as Type Ia.
  In the NIR, on the other hand, their spectral features are
  drastically different, revealing possible separate origins.  The
  optical spectra of SNe 1994D \citep{1997ASIC..486.....R}, 2009dc
  \citep{2011MNRAS.412.2735T}, and 2005hk \citep{2007PASP..119..360P}
  were previously published.  The GNIRS spectrum of ASASSN-15hy and
  FIRE spectra of SNe 2012fr and 2013gr were obtained as part of the
  CSP-II NIR spectroscopy program.}
\label{f:iapec}
\end{figure*}

While the primary goal of the CSP-II NIR spectroscopy data set is to
characterize SN~Ia K-corrections in the NIR, the same data set is also
a valuable resource for studying the physics and the progenitors of
SNe~Ia.  The NIR spectral region has several unsaturated, strong and
relatively isolated lines.  Furthermore, optical and NIR spectra at
the same phase probe completely different regions in the ejecta
\citep{1998ApJ...496..908W}.  Figure~\ref{f:iapec} shows comparison of
spectral features of normal and peculiar \sneia\ in the SN 2002cx-like
and the ``super-Chandrasekhar'' subtypes.  In the optical, the
spectral features between normal and peculiar \sneia\ are broadly
similar.  Indeed, that is why they are all classified as Type Ia. In
the NIR, on the other hand, their spectral features are drastically
different, revealing possible separate origins.  Recent advances in
the modeling of this wavelength region are also rapidly expanding the
NIR tool set.  We summarize some of these diagnostics here.  More
detailed analysis for each will be presented in forthcoming papers.

\subsection{Unburned Material via \ci\ \lam1.0693 \um}
\label{ss:carbon}

%explosion mechanisms

The general consensus for the origin of a SN~Ia is the thermonuclear
explosion of a carbon-oxygen white dwarf \citep{1960ApJ...132..565H}.
Since oxygen is also produced from carbon burning, carbon provides the
most direct probe of the primordial material from the progenitor white
dwarf.  The quantity, distribution and incidence of unburned carbon
provides important constraints for explosion models.  The
Chandrasekhar-mass DDT scenario predicts nearly complete carbon
burning for normal \sneia\ \citep{2009Natur.460..869K} and increasing
amounts of surviving carbon for fainter
\sneia\ \citep{2002ApJ...568..791H}.  The pulsating class of DDT
models also leaves large amounts of unburned carbon, even for normal
objects \citep{1995ApJ...444..831H}.  On the other hand, unburned
carbon is not expected to survive in the explosions of
sub-Chandrasekhar mass white dwarfs, except for a small amount below
the outer layer of iron-group elements produced during the surface
helium detonation \citep[e.g.,][]{2010A&A...514A..53F}.

%optical studies

Until recently, the study of unburned material had mainly relied on
the optical \cii\ \lam0.6580 \um\ line, which is marred by several
selection biases.  It is a weak spectral feature situated near the
emission component of the strong \siii\ \lam0.6355 \um\ P-Cygni
profile, such that high-velocity carbon may be buried in the silicon
absorption and measured velocities suffer from limb variation effects
\citep{1990A&A...229..191H}.  It also tends to fade rapidly after
explosion, which makes obtaining a complete sample rather difficult.
Surveys of large optical spectroscopic samples found that
$20-30$\%\ of pre-maximum spectra show \cii\ signatures
\citep{2011ApJ...743...27T, 2011ApJ...732...30P, 2012ApJ...745...74F,
  2012MNRAS.425.1917S}.

%advantages of NIR

\begin{figure*}
\centering
\includegraphics[width=1.0\textwidth,clip=true]{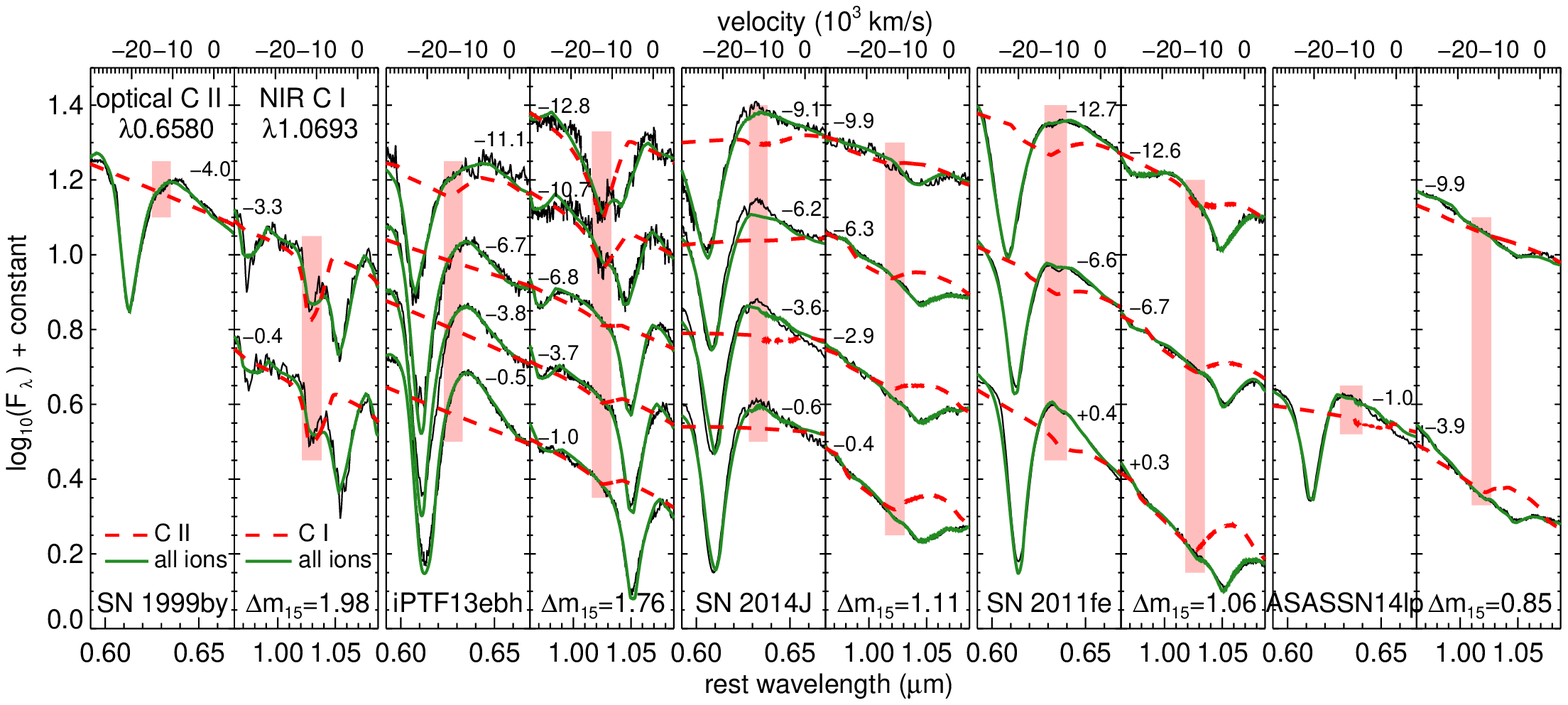}
\caption{Comparison between the optical \cii\ \lam0.6580 \um\ and the
  NIR \ci\ \lam1.0693 \um\ lines of the five SNe~Ia with
  \ci\ detections.  This plot is an expansion of Figure~10 in
  \citet{2015A&A...578A...9H} with the addition of the slow-declining
  ASASSN-14lp.  The optical/NIR pair is selected such that the spectra
  are close in phase for each SN~Ia.  The phases relative to $B$
  maximum are labeled.  The velocity axes are plotted with respect to
  each carbon line.  The \texttt{SYNAPPS} \citep{2011PASP..123..237T}
  fits are plotted in green solid curves.  We also plot the isolated
  influence from the carbon lines of the same fits in red dashed
  curves.  The vertical pink bars provide guides to the locations of
  the carbon absorption troughs, or approximately where they would be
  if the features are indeed present.  Except at the very early epoch
  of SN~2011fe, the NIR \ci\ line is always stronger than the optical
  \cii\ line.}
\label{f:carbon}
\end{figure*}

In contrast to the weak and fast-fading nature of the optical
\cii\ \lam0.6580 \um\ line, the NIR \ci\ \lam1.0693 \um\ line may be
much stronger.  Also, it increases in strength toward maximum light
for normal-bright objects \citep{2013ApJ...766...72H}.  The delayed
onset of the NIR \ci\ line may be a manifestation of the recombination
from \cii\ to \ci\ as the ejecta expands and cools, and highlights the
potential of the NIR \ci\ line to secure more representative
properties of unburned material in SNe~Ia.  With the aid of
\texttt{SYNAPPS} \citep{2011PASP..123..237T}, the NIR \ci\ feature has
been identified in several SNe~Ia with a large range of peak
magnitudes \citep[Figure~\ref{f:carbon};][]{2013ApJ...766...72H,
  2015A&A...578A...9H, 2015ApJ...798...39M}.  While the current number
of NIR carbon detections is small, the incidence of carbon appears to
be ubiquitous, with the fast-declining, fainter \sneia\ having the
strongest \ci\ lines.  This preliminary result appears to contradict
the claim of \citet{2017MNRAS.470..157B} that fast-declining faint
\sneia\ result from the explosions of sub-Chandrasekhar-mass white
dwarfs.  However, more detailed analysis is needed to determine
whether the small amount of unburned carbon beneath the surface
iron-peak layer in sub-Chandrasekhar-mass models can produce the
observed carbon features.  With the CSP-II sample of $\sim120$
pre-maximum NIR spectra of $\sim50$ SNe~Ia, we aim to characterize the
properties of unburned material in SNe~Ia in an unbiased fashion.

%unburned helium

In the sub-Chandrasekhar helium detonation scenario, unburned helium
left over from the surface helium layer can potentially produce strong
NIR \hei\ \lam1.0830 and \lam2.0581 \um\ lines, as we have shown for
other supernova types in Figure~\ref{f:type}.
\citet{2017A&A...599A..46B} explored these features, and the
possibility of confusion with the \ci\ \lam1.0693 \um\ line.  This
emphasizes the importance of identifying multiple lines of the same
ion when searching for unburned carbon, as was done for SN~1999by
\citep{2002ApJ...568..791H} and iPTF13ebh \citep{2015A&A...578A...9H}.
Note that both of these objects are fast-declining, fainter \sneia,
which tend to show the strongest \ci\ lines.

\subsection{Radioactive Nickel via $H$-band Break}
\label{ss:hbreak}

A dramatic shift in the amount of line blanketing from iron-group
elements takes place near 1.5 \um\ and produces the strong $H$-band
break after maximum light \citep{1998ApJ...496..908W}.
Observationally, SNe~Ia are found to have uniform evolution of the
$H$-band features.  The $H$-band break appears at $\sim3$ days past
maximum and rapidly increases in strength as the feature formed by
iron-peak elements becomes fully exposed at $\sim12$ days past
maximum.  It then steadily decreases in strength at very similar rates
for all SNe~Ia out to approximately one month past maximum.  The
maximum size of the $H$-band break, when the iron-peak complex is
fully exposed, is also found to correlate with the SN~Ia light curve
decline rate \dm\ \citep{2013ApJ...766...72H}.

The size and velocity shift of the $H$-band break probes the amount
and distribution of radioactive \nirad, and the time evolution of the
$H$-band break is related to the amount of intermediate-mass elements
acting as the ``shielding mass.''  In the Chandrasekhar-mass DDT
scenario, for example, there is an increasing amount of
intermediate-mass elements produced with decreasing
\nirad\ production; while for the sub-Chandrasekhar-mass He detonation
scenario, the \nirad\ production increases with the total mass.  Each
scenario therefore produces a unique rate at which the $H$-band
complex is exposed and a specific correlation between the strength of
the $H$-band break and \dm.  CSP-II has gathered $\sim300$ NIR spectra
of $\sim120$ SNe~Ia between maximum light and $+30$ days.  We aim to
use this sample to confirm the strong correlation with decline rate
discovered by \citet{2013ApJ...766...72H, 2015A&A...578A...9H} with
only ten SNe~Ia, in addition to improving the measurements of the rate
at which the $H$-band complex is exposed, which is currently poorly
constrained.

\subsection{Companion Signature via \pbeta\ and \hei}
\label{ss:companion}

The nature of the companion star in the progenitor system is still a
mystery.  The collision between the SN~Ia ejecta and the companion may
produce detectable signatures that can be used to place constraints on
the properties of the companion star
\citep[e.g.,][]{2010ApJ...708.1025K}, although predictions on whether
the signature is detectable may be model-dependent
\citep{1996MNRAS.283.1355C}.  In particular, if the companion star has
a hydrogen or helium-rich envelope in a single-degenerate system, the
envelope is expected to be stripped away during the collision and be
embedded deep within the low-velocity part of the expanding SN~Ia
ejecta \citep{2000ApJS..128..615M}.  This has been suggested to
produce \halpha\ in late-time spectra, beyond 200 days.  So far, none
has been detected in nearby SNe~Ia
\citep[e.g.,][]{2005A&A...443..649M, 2013ApJ...762L...5S,
  2018ApJ...863...24S}.  More recently, \citet{2014ApJ...794...37M}
suggested that NIR \pbeta\ companion signature is potentially much
stronger than \halpha, and appears much earlier.  The strong NIR
\hei\ lines can also be easily detected \citep{2018ApJ...852L...6B}.
These desirable properties could lead to the first detection of
embedded hydrogen or helium stripped from the companion star or at
least place stronger constraints on its absence.  We published the
first searches for the \pbeta\ emission in the NIR.  Although these
were also non-detections, they placed upper limits of $<0.1$
$M_{\sun}$ of stripped hydrogen from the companion stars of SNe 2014J
\citep{2016ApJ...822L..16S} and 2017cbv \citep{2017ApJ...845L..11H}.

Two main challenges for the detection of these so-far elusive hydrogen
lines are: 1) viewing angle effects and 2) diversity of the SN~Ia
feature underneath the hydrogen emissions, both requiring a
statistically significant sample.  The CSP-II data set with $\sim100$
NIR spectra of $\sim50$ SNe~Ia between one and two months past maximum
currently is the best data set for such a study.

\subsection{Neutron Content via \nifor\ 1.939 \um}
\label{ss:neutron}

Recent theoretical studies of \sneia\ undergoing the transition from
the photospheric to nebular phase identified a strong NIR emission
feature as \nifor\ $\lambda$1.939 $\mu$m \citep{2014ApJ...792..120F,
  2018MNRAS.474.3187W}, although this identification had been disputed
by \citet{2015MNRAS.448.2766B}.  At such late phases, all of the
radioactive \nirad\ has decayed into \corad.  If the observed line is
indeed attributed to \nifor, it must then be produced by the stable
isotope \nista.  Furthermore, since we are likely probing the neutron
content in the inner region at these late phases, the detection of
\nifor\ features would indicate high-density burning
\citep{1986A&A...158...17T, 1999ApJS..125..439I} and may be less
sensitive to the metallicity of the progenitor
\citep{1998ApJ...495..617H, 2003ApJ...590L..83T} or the neutronization
in the simmering phase \citep{2008ApJ...673.1009P}.  High-density
burning is a hallmark of Chandrasekhar-mass DDT models, while
sub-Chandrasekhar explosion models tend to underproduce stable
neutron-rich elements \citep{2013A&A...559L...5S}.

Spectra in the NIR obtained during the transition to nebular phase are
still quite rare.  Furthermore, the \nifor\ $\lambda$1.939 $\mu$m line
is close to the strong telluric water vapor absorptions between $H$
and $K$ bands.  \citet{2014ApJ...792..120F} gathered four NIR spectra
from the literature and two new observations of SN~2014J between 50
and 100 days past explosion.  Only the two NIR spectra of SN~2014J had
adequate wavelength coverage, telluric correction, and S/N ratio to
detect the presence of this feature.  The feature was also reported in
the NIR nebular spectra of SN~2014J obtained by
\citet{2018arXiv180502420D}.  With the CSP-II sample of $\sim100$
transitional to nebular phase NIR spectra of $\sim40$ SNe~Ia, we aim
to analyze the diversity in the properties of this feature.
Figure~\ref{f:12fr} shows an example of our time-series data, taken of
SN~2012fr.  The feature reported by \citet{2014ApJ...792..120F} was
indeed also present in SN~2012fr and evolved to be its strongest
feature in the $K$ band during the transitional phase (+82 days past
$B$ maximum in Figure~\ref{f:12fr}).  By the nebular phase, the
feature became much weaker, and our nebular spectrum at +367 days
unfortunately did not have a high enough S/N ratio to confirm the
presence of this feature.  Further study on the modeling side is also
planned to secure the identification of this line.

\subsection{Central Density and Magnetic Field via \fefor\ 1.644 \um}

Nebular phase observations of line profiles can reveal the isotopic
structure of the inner region.  In particular, the forbidden
\fefor\ 1.644 \um\ line is ideal for such studies, since it is strong
and unblended.  It can be used to measure the properties of the
progenitor system and explosion, such as asphericity, initial central
density, and magnetic field \citep{2014ApJ...795...84P,
  2015ApJ...806..107D, 2018ApJ...861..119D}.  Up to 200 days past
explosion, the line profile can be used to analyze the overall
chemical and density structure of the exploding white dwarf without
considering the magnetic field, since the energy deposition by gamma
ray photons dominates.  Measurements of the central density can lead
us to the accretion history of the pre-explosion white dwarf
\citep{1984ApJ...286..644N} and the amount of stable \nista\ produced
\citep{2006NewAR..50..470H}.  Note that updated electron capture rates
\citep[e.g.,][]{2000ApJ...536..934B} no longer yield enough stable
iron group elements to produce the ``flat-top'' profiles emphasized by
\citet{2018MNRAS.477.3567M}, except in the most extreme cases.
Rather, increasing central density widens the line width of the
\fefor\ 1.644 \um\ line and the effects can be measured with moderate
S/N ratio nebular spectra \citep{2015ApJ...806..107D,
  2018ApJ...861..119D}.  Going to even later phases, beyond one year
past explosion, the time evolution of the line profile is sensitive to
positron transport effects.  The escape probability of the positrons,
which directly influences the width of the \fefor\ line, depends
strongly on the size and the morphology of the magnetic fields.

For the Chandrasekhar-mass DDT scenario, 3D hydrodynamical simulations
consistently predict large-scale mixing of the ejecta due to
Rayleigh-Taylor instabilities during the deflagration burning phase
\citep{2006A&A...453..203R}, producing results which are inconsistent
with observations.  Recently, \citet{2018ApJ...858...13H} found that
the magnetic field plays an important role during this burning phase.
Their magnetohydrodynamical study showed increasing suppression of the
instabilities starting at $\sim10^9$ G.  Such magnetic fields may be
generated during the pre-explosive carbon-simmering phase of
Chandrasekhar-mass explosions \citep{2008ApJ...678.1158P}.  Current
measurements based on a handful of NIR nebular spectra of nearby
SNe~Ia indicated relatively high central densities and high magnetic
fields \citep[e.g.,][]{2014ApJ...795...84P, 2015ApJ...806..107D}.  We
are extending the analysis to the CSP-II sample of $\sim20$ NIR
spectra of seven SNe~Ia at phases beyond 200 days past explosion.
Meanwhile, ongoing programs at Gemini and LCO, as part of CSP-II,
continue to monitor the time evolution of the \fefor\ line profile of
nearby SNe~Ia.

%%%%%%%%%%%%%%%%
%% Conclusion %%
%%%%%%%%%%%%%%%%

\section{Summary}
\label{s:conclusion}

This paper presents an introduction to the CSP-II NIR spectroscopy
program, which we carried out from 2011 to 2015.  This project
represents an important step for any current and future
\snia\ cosmological experiments based on the rest-frame NIR.  These
experiments have been shown to reduce errors by effectively
circumventing the uncertain dust laws and empirical width-luminosity
relation.  The data set collected is unique in its large sample size,
time-series observations, complementary light-curve and optical
spectroscopic observations, and improvement in the telluric regions.
Such a sample makes it possible to quantify the representative NIR
spectroscopic time evolution and diversity of \sneia, allowing NIR
cosmological experiments to reach their true potential.  By exploring
the NIR wavelength window, many diagnostics for the explosion
mechanisms and the progenitor systems of \sneia\ also become
available.  Understanding the origins of these events represents a
second and independent way to reduce systematic errors in
\snia\ cosmology.  While our focus is on normal \sneia\ for
cosmological studies, comparisons of NIR spectroscopic features
between normal and peculiar \sneia\ (Figure~\ref{f:iapec}), and
between supernovae of various types (Figure~\ref{f:type}), highlight
the identifying features and offer clues to the origins of these
events.

\acknowledgements We are pleased to acknowledge the following
individuals (in alphabetical order) for their assistance in obtaining
the NIR spectroscopic data set: Y.~Beletsky, G.~Blanc, T.~Dupuy,
N.~Elias-Rosa, R.~Foley, L.~W.~Hsiao, B.~Madore, A.~Monson, E.~Newton,
D.~Osip, P.~Palunas, J.~L.~Prieto, M.~Rouch, S.~Schulze, and
M.~Turatto.  We also thank the Las Campanas technical staff for their
continued support over the years.  The CSP-II has been supported by
NSF grants AST-1008343, AST-1613426, AST-1613455, and AST-1613472, as
well as, the Danish Agency for Science and Technology and Innovation
through a Sapere Aude Level 2 grant.  M.~S. acknowledges funding by a
research grant (13261) from VILLUM FONDEN.  Research by D.~J.~S. is
supported by NSF grants AST-1821967, 1821987, 1813708, and 1813466.
N.~B.~S. and K.~K. acknowledge support from the The George P. and
Cynthia Woods Mitchell Institute for Fundamental Physics and
Astronomy.  T.~D. is supported by an appointment to the NASA
Postdoctoral Program at the Goddard Space Flight Center, administered
by Universities Space Research Association under contract with NASA.
L.~W. acknowledges the support by the Chinese Academy of Sciences
(CAS), through a grant to the CAS South America Center for Astronomy
(CASSACA) in Santiago, Chile.  The bulk of the data presented here was
obtained with the 1 m Swope, 2.5 m du Pont, and the 6.5 m Magellan
Telescopes at the Las Campanas Observatory.  Observations were also
obtained at the Gemini Observatory (program ID: GN-2011B-Q-68,
GN-2012A-Q-59, GN-2012A-Q-69, GN-2013B-Q-76, GN-2014A-Q-52,
GN-2014B-Q-13, GN-2014B-Q-70, GN-2015B-Q-7, GS-2015B-Q-5), which is
operated by the Association of Universities for Research in Astronomy,
Inc., under a cooperative agreement with the NSF on behalf of the
Gemini partnership: the National Science Foundation (United States),
the National Research Council (Canada), CONICYT (Chile), Ministerio de
Ciencia, Tecnolog\'{i}a e Innovaci\'{o}n Productiva (Argentina), and
Minist\'{e}rio da Ci\^{e}ncia, Tecnologia e Inova\c{c}\~{a}o (Brazil).
Also based on observations collected at the European Organisation for
Astronomical Research in the Southern Hemisphere under ESO programme
088.D-0222.  This research used resources of the National Energy
Research Scientific Computing Center (NERSC), a U.S. Department of
Energy Office of Science User Facility operated under Contract
No. DE-AC02-05CH11231.  We have also made use of the NASA/IPAC
Extragalactic Database (NED) which is operated by the Jet Propulsion
Laboratory, California Institute of Technology, under contract with
the National Aeronautics and Space Administration.

\facilities{Magellan Baade (FIRE), du~Pont (WFCCD) Gemini North (GNIRS
  near-infrared spectrograph), Gemini South (FLAMINGOS-2), VLT
  (ISAAC), IRTF (SpeX), NOT (ALFOSC), La Silla-QUEST, CRTS, PTF, iPTF,
  OGLE, ASAS-SN, PS1, KISS, ISSP, MASTER, SMT}

\software{
\texttt{firehose} \citep{2013PASP..125..270S},
\texttt{GELATO} \citep{2008A&A...488..383H},
\texttt{SNID} \citep{2007ApJ...666.1024B},
\texttt{Spextool} \citep{2004PASP..116..362C},
\texttt{superfit} \citep{2005ApJ...634.1190H},
\texttt{SYNAPPS} \citep{2011PASP..123..237T},
\texttt{XDGNIRS},
\texttt{xtellcor} \citep{2003PASP..115..389V}
}

\startlongtable
\begin{deluxetable*}{lccll}
\tabletypesize{\scriptsize}
\tablecolumns{5}
\tablecaption{Summary of CSP-II NIR spectroscopy of \sneia\label{t:nirspec_name}}
\tablehead{
\colhead{SN Name} & 
\colhead{Number of NIR spectra} & 
\colhead{Sample\tablenotemark{a}} & 
\colhead{$z_{\rm{helio}}$\tablenotemark{b}} &
\colhead{Classification by FIRE\tablenotemark{c}}}
\startdata
       ASASSN-13ar (SN~2013dl) &   3 &      &  0.0178 &  \\
                   ASASSN-13av &   3 &      &  0.0173 &  \\
                   ASASSN-14ad &   6 &    P &  0.0264 &  \\
                   ASASSN-14eu &   1 &      &  0.0227 &  \\
                   ASASSN-14hp &   1 &    C &  0.0389 &  \\
                   ASASSN-14hu &   2 &    P &  0.0216 &  \\
                   ASASSN-14jc &   1 &    P &  0.0113 &  \\
                   ASASSN-14jg &   4 &    P &  0.0148 &  \\
                   ASASSN-14lo &   1 &    P &  0.0199 &  \\
                   ASASSN-14lp &  25 &    P &  0.0051 &  \\
                   ASASSN-14lq &   1 &    P &  0.0262 &  \\
                   ASASSN-14lw &   3 &    P &  0.0209 & ATel 6832 \\
                   ASASSN-14me &   3 &    P &  0.018\tablenotemark{d} &  \\
                   ASASSN-14mw &   3 &  C,P &  0.0274 &  \\
                   ASASSN-14my &   4 &    P &  0.0205 &  \\
                   ASASSN-15aj &   2 &    P &  0.0109 &  \\
                   ASASSN-15as &   2 &  C,P &  0.0286 & ATel 6920 \\
                   ASASSN-15ba &   2 &    P &  0.0231 &  \\
                   ASASSN-15be &   2 &    P &  0.0219 &  \\
                   ASASSN-15bm &   3 &    P &  0.0208 &  \\
                   ASASSN-15cd &   1 &    C &  0.0344 &  \\
                   ASASSN-15eb &   1 &    P &  0.0165 &  \\
                   ASASSN-15fr &   1 &  C,P &  0.0334 &  \\
                   ASASSN-15fy &   1 &      &  0.025\tablenotemark{d} & ATel 7354 \\
                   ASASSN-15ga &   3 &    P &  0.0066 &  \\
                   ASASSN-15go &   1 &    P &  0.0189 &  \\
                   ASASSN-15gr &   1 &    P &  0.0243 &  \\
                   ASASSN-15hf &   3 &    P &  0.0062 &  \\
                   ASASSN-15hx &   6 &    P &  0.008\tablenotemark{d} &  \\
                   ASASSN-15hy &   6 &      &  0.025\tablenotemark{d} &  \\
\hline
       CSS110414:145909+071804 (SN~2011cg) &   1 &      &  0.020\tablenotemark{d} &  \\
       CSS110504:101800-023241 (SN~2011ci) &   1 &      &  0.0472 &  \\
       CSS110604:130707-011044 (SN~2011dj) &   1 &      &  0.0185 &  \\
       CSS111231:145323+025743 (SN~2011jt) &   1 &    P &  0.0278 &  \\
       CSS120224:145405+220541 (SN~2012aq) &   1 &    C &  0.052\tablenotemark{d} &  \\
       CSS120301:162036-102738 (SN~2012ar) &   6 &  C,P &  0.0283 & CBET 3041 \\
       CSS121114:090202+101800 &   1 &    C &  0.0371 &  \\
       SSS130304:114445-203141 (SN~2013ao) &   6 &      &  0.0435 &  \\
       CSS130315:114144-171348 &   1 &      &  0.050\tablenotemark{d} &  \\
       CSS130614:233746+144237 (SN~2013dn) &   3 &      &  0.0562 &  \\
       CSS131225:030144-092044 &   1 &      &  0.040\tablenotemark{d} & ATel 5702 \\
       CSS150214:140955+173155 (SN~2015bo) &   1 &    P &  0.0162 &  \\
         SNhunt37  (SN~2011ae) &   1 &      &  0.0060 &  \\
         SNhunt46  (SN~2011be) &   1 &      &  0.0340 &  \\
         SNhunt177 (SN~2013az) &   1 &    C &  0.0373 & CBET 3457, ATel 4935 \\
         SNhunt178 (SN~2013bc) &   1 &    P &  0.0225 & CBET 3468, ATel 4948 \\
         SNhunt188 (SN~2013bz) &   1 &    P &  0.0192 &  \\
         SNhunt224 (SN~2013hs) &   1 &      &  0.0195 & CBET 3770, ATel 5702 \\
         SNhunt229 (SN~2014D)  &   2 &    P &  0.0082 & CBET 3778 \\
         SNhunt281 (SN~2015bp) &   9 &    P &  0.0041 &  \\
\hline
            KISS15n (SN~2015M) &   1 &    P &  0.0231\tablenotemark{e} &  \\
\hline
                       LSQ11bk &   2 &    C &  0.0403 &  \\
                       LSQ11ot &   3 &  C,P &  0.0273 &  \\
                       LSQ11pn &   1 &  C,P &  0.0327 &  \\
                      LSQ12bia &   1 &      &  0.053\tablenotemark{d} &  \\
                      LSQ12btn &   2 &    C &  0.0542 &  \\
                      LSQ12cpf &   2 &      &  0.0286 &  \\
                      LSQ12dbr &   1 &      &  0.020\tablenotemark{d} &  \\
                      LSQ12frx &   1 &      &  0.030\tablenotemark{d} &  \\
                      LSQ12fuk &   2 &    P &  0.0206 &  \\
                      LSQ12fxd &   7 &  C,P &  0.0312 &  \\
                      LSQ12gdj &   5 &  C,P &  0.0303 &  \\
                      LSQ12hzj &   1 &  C,P &  0.0334 &  \\
                       LSQ13pf &   1 &    C &  0.0861 & ATel 4916 \\
                       LSQ13ry &   2 &  C,P &  0.0299 & ATel 4935 \\
          LSQ13aiz (SN~2013cs) &   3 &    P &  0.0092 &  \\
                      LSQ13dsm &   3 &    C &  0.0424 & ATel 5714 \\
                       LSQ14xi &   1 &    C &  0.0508 &  \\
          LSQ14ajn (SN~2014ah) &   2 &    P &  0.0210 &  \\
                      LSQ14dsu &   1 &      &  0.0196 &  \\
                      LSQ15aae &   1 &    C &  0.0516 &  \\
                      LSQ15adm &   1 &      &  0.0723 &  \\
\hline
 MASTER OT J030559.89+043238.2 &   3 &    P &  0.0282 &  \\
\hline
              OGLE-2014-SN-019 &   1 &    C &  0.0359 &  \\
\hline
                      PS1-14ra &   3 &  C,P &  0.0281 &  \\
                        PS15sv &   3 &  C,P &  0.0333 &  \\
\hline
         PSN J13471211-2422171 &   1 &    P &  0.0199 &  \\
\hline
                      PTF11bju &   1 &      &  0.0323 &  \\
          PTF11kly (SN~2011fe) &  15 &      &  0.0008 &  \\
          PTF11pbp (SN~2011hb) &   3 &  C,P &  0.0289 &  \\
                      PTF11qnr &   3 &    P &  0.0162 &  \\
                      PTF12ena &   1 &      &  0.0166 &  \\
                     iPTF13asv &   1 &      &  0.035\tablenotemark{d} &  \\
                     iPTF13dge &   1 &      &  0.0159 &  \\
                     iPTF13duj &   4 &    P &  0.0170 &  \\
                     iPTF13dym &   1 &    C &  0.0422 &  \\
                     iPTF13dzm &   1 &      &  0.0158 &  \\
                     iPTF13ebh &  15 &    P &  0.0133 & ATel 5580 \\
                     iPTF14aaf &   1 &      &  0.0589 &  \\
                     iPTF14abh &   1 &      &  0.0237 &  \\
                     iPTF14ans &   1 &      &  0.0318 &  \\
                     iPTF14bdn &   1 &      &  0.0156 &  \\
                      iPTF14sz &   1 &      &  0.0288 &  \\
                       iPTF14w &   7 &    P &  0.0189 &  \\
          iPTF14yw (SN~2014aa) &   4 &    P &  0.0170 &  \\
                      iPTF14yy &   1 &    C &  0.0431 &  \\
         iPTF14fpg (SN~2014dk) &   4 &  C,P &  0.0319 &  \\
                      iPTF15ku &   5 &      &  0.0282 &  \\
\hline
ROTSE3 J123935.1+163512 (SN~2012G) &   2 &    P &  0.0266 &  \\
\hline
                     SN~2011at &   2 &      &  0.0068 &  \\
                     SN~2011bf &   2 &      &  0.0167 &  \\
                     SN~2011ce &   3 &      &  0.0086 &  \\
                     SN~2011di &   1 &      &  0.0147 &  \\
                     SN~2011iv &  20 &    P &  0.0065 &  \\
                     SN~2011iy &  13 &    P &  0.0043 &  \\
                     SN~2011jh &  11 &    P &  0.0078 &  \\
                     SN~2011jl &   2 &      &  0.0100 &  \\
                      SN~2012E &   4 &    P &  0.0203 &  \\
                      SN~2012U &   1 &    P &  0.0197 &  \\
                      SN~2012Z &  12 &      &  0.0071 &  \\
                     SN~2012ah &   1 &    P &  0.0124 &  \\
                     SN~2012bl &   9 &    P &  0.0187 &  \\
                     SN~2012bo &   7 &    P &  0.0254 &  \\
                     SN~2012cg &  19 &      &  0.0015 & ATel 4119 \\
                     SN~2012db &   1 &      &  0.0193 &  \\
                     SN~2012fr &  40 &    P &  0.0055 &  \\
                     SN~2012gm &   2 &    P &  0.0148 &  \\
                     SN~2012hd &   5 &    P &  0.0120 &  \\
                     SN~2012hr &  12 &    P &  0.0076 & CBET 3346, ATel 4663 \\
                     SN~2012ht &  17 &    P &  0.0036 &  \\
                     SN~2012id &   3 &    P &  0.0157 &  \\
                     SN~2012ij &   3 &    P &  0.0110 &  \\
                      SN~2013E &  12 &    P &  0.0094 &  \\
                      SN~2013H &   4 &    P &  0.0155 &  \\
                      SN~2013M &   3 &  C,P &  0.0350 &  \\
                      SN~2013N &   1 &      &  0.0256 &  \\
                      SN~2013U &   2 &  C,P &  0.0345 &  \\
                     SN~2013aa &  16 &    P &  0.0040 &  \\
                     SN~2013aj &   9 &    P &  0.0091 &  \\
                     SN~2013ay &   2 &    P &  0.0157 & CBET 3456, ATel 4935 \\
                     SN~2013cg &   1 &    P &  0.0080 &  \\
                     SN~2013ct &   7 &    P &  0.0038 & CBET 3539, ATel 5081 \\
                     SN~2013dh &   2 &      &  0.0134 &  \\
                     SN~2013dt &   3 &      &  0.0241 &  \\
                     SN~2013fb &   1 &      &  0.0170 &  \\
                     SN~2013fy &   6 &  C,P &  0.0309 &  \\
                     SN~2013fz &   7 &    P &  0.0206 &  \\
                     SN~2013gr &  12 &      &  0.0074 & CBET 3733, ATel 5612 \\
                     SN~2013gv &   4 &  C,P &  0.0341 &  \\
                     SN~2013gy &  16 &    P &  0.0140 &  \\
                     SN~2013hh &   7 &    P &  0.0130 &  \\
                     SN~2013hl &   1 &      &  0.0262 & CBET 3759, ATel 5664 \\
                     SN~2013hn &   4 &    P &  0.0151 &  \\
                      SN~2014I &   5 &  C,P &  0.0300 &  \\
                      SN~2014J &  49 &      &  0.0007 &  \\
                      SN~2014Z &   1 &    P &  0.0213 & CBET 3822, ATel 5959 \\
                     SN~2014ao &   5 &    P &  0.0141 &  \\
                     SN~2014at &   2 &  C,P &  0.0322 &  \\
                     SN~2014ba &   2 &    P &  0.0058 &  \\
                     SN~2014cd &   1 &      &  0.0206 &  \\
                     SN~2014ck &   7 &      &  0.0050 &  \\
                     SN~2014cz &   1 &      &  0.0260 & CBET 3965, ATel 6442 \\
                     SN~2014du &   1 &  C,P &  0.0325 &  \\
                     SN~2014eg &   1 &    P &  0.0186 &  \\
                     SN~2014ek &   1 &      &  0.0231 &  \\
                      SN~2015F &   6 &    P &  0.0049 &  \\
                      SN~2015H &   3 &      &  0.0125 &  \\
\enddata
\tablenotetext{a}{``C'' and ``P'' indicate that the \snia\ is in the ``Cosmology'' and ``Physics'' samples, respectively.}
\tablenotetext{b}{Host redshifts are from the NASA/IPAC Extragalactic Database (NED) or measured by CSP-II, unless otherwise noted.}
\tablenotetext{c}{The column lists the ATel and CBET numbers for the \sneia\ classified by FIRE.}
\tablenotetext{d}{The redshift is derived from the SN spectrum.}
\tablenotetext{e}{The redshift of Coma Cluster is adopted.}
%\tablecomments{}
\end{deluxetable*}

\begin{deluxetable*}{llcccccccc}
\tablecolumns{10}
\tablecaption{Number of NIR spectra by instrument and SN type\label{t:nirspec}}
\tablehead{
\colhead{Telecope} & 
\colhead{Instrument} & 
\colhead{Ia} & 
\colhead{2002cx-like} & 
\colhead{Ibc} & 
\colhead{II} & 
\colhead{IIn} & 
\colhead{SLSN} & 
\colhead{Others} & 
\colhead{Total}}
\startdata
Magellan Baade & FIRE       & 450 &  27 &  97 &  79 &  35 &   9 &   4 & 701 \\
Magellan Clay  & MMIRS      &   7 &   1 &   2 &   1 &   0 &   0 &   1 &  12 \\
Gemini North   & GNIRS      &  63 &   7 &   0 &   0 &   0 &   0 &   0 &  70 \\
Gemini South   & FLAMINGOS2 &  17 &   0 &   0 &   1 &   0 &   0 &   0 &  18 \\
IRTF           & SpeX       &  40 &   1 &   6 &   9 &   2 &   0 &   0 &  58 \\
VLT            & ISAAC      &   8 &   4 &   0 &   0 &   0 &   0 &   0 &  12 \\
NTT            & SofI       &   4 &   1 &   2 &   0 &   0 &   0 &   0 &   7 \\
Mt Abu         & NICS       &  31 &   0 &   0 &   0 &   0 &   0 &   0 &  31 \\
\hline
total spectra  &            & 620 &  41 & 107 &  90 &  37 &   9 &   5 & 909 \\
total SNe      &            & 149 &   8 &  42 &  31 &  12 &   5 &   2 & 249
\enddata
%\tablecomments{}
\end{deluxetable*}

\end{CJK*}

\begin{thebibliography}{}
\bibitem[Baltay et al.(2013)]{2013PASP..125..683B} Baltay, C., et al.\ 
2013, Publications of the Astronomical Society of the Pacific, 125, 683 
\bibitem[Barone-Nugent et al.(2012)]{2012MNRAS.425.1007B} Barone-Nugent, 
R.~L., et al.\ 2012, Monthly Notices of the Royal Astronomical Society, 
425, 1007 
\bibitem[Betoule et 
al.(2014)]{2014A&A...568A..22B} Betoule, M., et al.\ 2014, Astronomy and Astrophysics, 568, A22 
\bibitem[Blondin et al.(2015)]{2015MNRAS.448.2766B} Blondin, S., Dessart, 
L., 
\& Hillier, D.~J.\ 2015, Monthly Notices of the Royal Astronomical Society, 448, 2766 
\bibitem[Blondin et al.(2017)]{2017MNRAS.470..157B} Blondin, S., et al.\ 
2017, Monthly Notices of the Royal Astronomical Society, 470, 157 
\bibitem[Blondin 
\& Tonry(2007)]{2007ApJ...666.1024B} Blondin, S., \& Tonry, J.~L.\ 2007, The Astrophysical Journal, 666, 1024 
\bibitem[Boldt et al.(2014)]{2014PASP..126..324B} Boldt, L.~N., et al.\ 
2014, Publications of the Astronomical Society of the Pacific, 126, 324 
\bibitem[Boty{\'a}nszki et al.(2018)]{2018ApJ...852L...6B} Boty{\'a}nszki, 
J., Kasen, D., \& Plewa, T.\ 2018, The Astrophysical Journal, 852, L6 
\bibitem[Boyle et 
al.(2017)]{2017A&A...599A..46B} Boyle, A., et al.\ 2017, Astronomy and Astrophysics, 599, A46 
\bibitem[Brachwitz et al.(2000)]{2000ApJ...536..934B} Brachwitz, F., et 
al.\ 2000, The Astrophysical Journal, 536, 934 
\bibitem[Bufano et al.(2014)]{2014MNRAS.439.1807B} Bufano, F., et al.\ 
2014, Monthly Notices of the Royal Astronomical Society, 439, 1807 
\bibitem[Burns et al.(2014)]{2014ApJ...789...32B} Burns, C.~R., et al.\ 
2014, The Astrophysical Journal, 789, 32 
\bibitem[Burns et al.(2011)]{2011AJ....141...19B} Burns, C.~R., et al.\ 
2011, The Astronomical Journal, 141, 19 
\bibitem[Burns et al.(2018)]{B18} Burns, C.~R., et al.\
2018, ArXiv e-prints, arXiv:1809.06381
\bibitem[Conley et al.(2011)]{2011ApJS..192....1C} Conley, A., et al.\ 
2011, The Astrophysical Journal Supplement Series, 192, 1 
\bibitem[Contreras et al.(2010)]{2010AJ....139..519C} Contreras, C., et 
al.\ 2010, The Astronomical Journal, 139, 519 
\bibitem[Contreras et al.(2018)]{2018ApJ...859...24C} Contreras, C., et 
al.\ 2018, The Astrophysical Journal, 859, 24 
\bibitem[Cumming et al.(1996)]{1996MNRAS.283.1355C} Cumming, R.~J., et al.\ 
1996, Monthly Notices of the Royal Astronomical Society, 283, 1355 
\bibitem[Cushing et al.(2004)]{2004PASP..116..362C} Cushing, M.~C., Vacca, 
W.~D., 
\& Rayner, J.~T.\ 2004, Publications of the Astronomical Society of the Pacific, 116, 362 
\bibitem[Dall'Ora et al.(2014)]{2014ApJ...787..139D} Dall'Ora, M., et al.\ 
2014, The Astrophysical Journal, 787, 139 
\bibitem[Dhawan et al.(2018)]{2018arXiv180502420D} Dhawan, S., et al.\ 
2018, ArXiv e-prints, arXiv:1805.02420 
\bibitem[Diamond et al.(2018)]{2018ApJ...861..119D} Diamond, T.~R., et al.\ 
2018, The Astrophysical Journal, 861, 119 
\bibitem[Diamond et al.(2015)]{2015ApJ...806..107D} Diamond, T.~R., 
Hoeflich, P., \& Gerardy, C.~L.\ 2015, The Astrophysical Journal, 806, 107 
\bibitem[Drout et al.(2016)]{2016ApJ...821...57D} Drout, M.~R., et al.\ 
2016, The Astrophysical Journal, 821, 57 
\bibitem[Eikenberry et al.(2008)]{2008SPIE.7014E..0VE} Eikenberry, S., et 
al.\ 2008, Ground-based and Airborne Instrumentation for Astronomy II, 
7014, 70140V 
\bibitem[Elias et al.(2006)]{2006SPIE.6269E..4CE} Elias, J.~H., et al.\ 
2006, Society of Photo-Optical Instrumentation Engineers (SPIE) Conference 
Series, 6269, 62694C 
\bibitem[Filippenko(1982)]{1982PASP...94..715F} Filippenko, A.~V.\ 1982, 
Publications of the Astronomical Society of the Pacific, 94, 715 
\bibitem[Fink et 
al.(2010)]{2010A&A...514A..53F} Fink, M., et al.\ 2010, Astronomy and Astrophysics, 514, A53 
\bibitem[Folatelli et al.(2010)]{2010AJ....139..120F} Folatelli, G., et 
al.\ 2010, The Astronomical Journal, 139, 120 
\bibitem[Folatelli et al.(2012)]{2012ApJ...745...74F} Folatelli, G., et 
al.\ 2012, The Astrophysical Journal, 745, 74 
\bibitem[Foley et al.(2013)]{2013ApJ...767...57F} Foley, R.~J., et al.\ 
2013, The Astrophysical Journal, 767, 57 
\bibitem[Fox et al.(2015)]{2015MNRAS.447..772F} Fox, O.~D., et al.\ 2015, 
Monthly Notices of the Royal Astronomical Society, 447, 772 
\bibitem[Freedman et al.(2009)]{2009ApJ...704.1036F} Freedman, W.~L., et 
al.\ 2009, The Astrophysical Journal, 704, 1036 
\bibitem[Friesen et al.(2014)]{2014ApJ...792..120F} Friesen, B., et al.\ 
2014, The Astrophysical Journal, 792, 120 
\bibitem[Gall et 
al.(2018)]{2018A&A...611A..58G} Gall, C., et al.\ 2018, Astronomy and Astrophysics, 611, A58 
\bibitem[Gonz{\'a}lez-Mart{\'{\i}}n et 
al.(2013)]{2013A&A...553A..35G} Gonz{\'a}lez-Mart{\'{\i}}n, O., et al.\ 2013, Astronomy and Astrophysics, 553, A35 
\bibitem[Hachinger et al.(2008)]{2008MNRAS.389.1087H} Hachinger, S., et 
al.\ 2008, Monthly Notices of the Royal Astronomical Society, 389, 1087 
\bibitem[Hamuy et al.(2006)]{2006PASP..118....2H} Hamuy, M., et al.\ 2006, 
Publications of the Astronomical Society of the Pacific, 118, 2 
\bibitem[Harutyunyan et 
al.(2008)]{2008A&A...488..383H} Harutyunyan, A.~H., et al.\ 2008, Astronomy and Astrophysics, 488, 383 

\bibitem[Hoeflich(1990)]{1990A&A...229..191H} Hoeflich, P.\ 1990, Astronomy and Astrophysics, 229, 191 
\bibitem[H{\"o}flich(1995)]{1995ApJ...443...89H} H{\"o}flich, P.\ 1995, The 
Astrophysical Journal, 443, 89 
\bibitem[Hoeflich et al.(1995)]{1995ApJ...444..831H} Hoeflich, P., 
Khokhlov, A.~M., 
\& Wheeler, J.~C.\ 1995, The Astrophysical Journal, 444, 831 
\bibitem[H{\"o}flich et al.(1998)]{1998ApJ...495..617H} H{\"o}flich, P., 
Wheeler, J.~C., 
\& Thielemann, F.~K.\ 1998, The Astrophysical Journal, 495, 617 
\bibitem[H{\"o}flich et al.(2002)]{2002ApJ...568..791H} H{\"o}flich, P., et 
al.\ 2002, The Astrophysical Journal, 568, 791 
\bibitem[H{\"o}flich et al.(2006)]{2006NewAR..50..470H} H{\"o}flich, P., et 
al.\ 2006, New Astronomy Reviews, 50, 470 
\bibitem[Hoeflich et al.(2017)]{2017ApJ...846...58H} Hoeflich, P., et al.\ 
2017, The Astrophysical Journal, 846, 58 
\bibitem[Horne(1986)]{1986PASP...98..609H} Horne, K.\ 1986, Publications of 
the Astronomical Society of the Pacific, 98, 609 
\bibitem[Hosseinzadeh et al.(2017)]{2017ApJ...845L..11H} Hosseinzadeh, G., 
et al.\ 2017, The Astrophysical Journal, 845, L11 
\bibitem[Howell et al.(2005)]{2005ApJ...634.1190H} Howell, D.~A., et al.\ 
2005, The Astrophysical Journal, 634, 1190 
\bibitem[Howell et al.(2006)]{2006Natur.443..308H} Howell, D.~A., et al.\ 
2006, Nature, 443, 308 
\bibitem[Hoyle 
\& Fowler(1960)]{1960ApJ...132..565H} Hoyle, F., \& Fowler, W.~A.\ 1960, The Astrophysical Journal, 132, 565 
\bibitem[Hristov et al.(2018)]{2018ApJ...858...13H} Hristov, B., et al.\ 
2018, The Astrophysical Journal, 858, 13 
\bibitem[Hsiao(2009)]{2009PhDT.......228H} Hsiao, E.~Y.\ 2009, 
Ph.D.~Thesis, University of Victoria
\bibitem[Hsiao et al.(2007)]{2007ApJ...663.1187H} Hsiao, E.~Y., et al.\ 
2007, The Astrophysical Journal, 663, 1187 
\bibitem[Hsiao et al.(2013)]{2013ApJ...766...72H} Hsiao, E.~Y., et al.\ 
2013, The Astrophysical Journal, 766, 72 
\bibitem[Hsiao et 
al.(2015)]{2015A&A...578A...9H} Hsiao, E.~Y., et al.\ 2015, Astronomy and Astrophysics, 578, A9 
\bibitem[Iben 
\& Tutukov(1984)]{1984ApJS...54..335I} Iben, I., Jr., \& Tutukov, A.~V.\ 1984, The Astrophysical Journal Supplement Series, 54, 335 
\bibitem[Inserra et al.(2014)]{2014MNRAS.437L..51I} Inserra, C., et al.\ 
2014, Monthly Notices of the Royal Astronomical Society, 437, L51 
\bibitem[Inserra et al.(2016)]{2016MNRAS.459.2721I} Inserra, C., et al.\ 
2016, Monthly Notices of the Royal Astronomical Society, 459, 2721 
\bibitem[Iwamoto et al.(1999)]{1999ApJS..125..439I} Iwamoto, K., et al.\ 
1999, The Astrophysical Journal Supplement Series, 125, 439 
\bibitem[Kasen et al.(2009)]{2009Natur.460..869K} Kasen, D., R{\"o}pke, 
F.~K., \& Woosley, S.~E.\ 2009, Nature, 460, 869 
\bibitem[Kasen(2006)]{2006ApJ...649..939K} Kasen, D.\ 2006, The 
Astrophysical Journal, 649, 939 
\bibitem[Kasen(2010)]{2010ApJ...708.1025K} Kasen, D.\ 2010, The 
Astrophysical Journal, 708, 1025 
\bibitem[Kattner et al.(2012)]{2012PASP..124..114K} Kattner, S., et al.\ 
2012, Publications of the Astronomical Society of the Pacific, 124, 114 
\bibitem[Kelson(2003)]{2003PASP..115..688K} Kelson, D.~D.\ 2003, 
Publications of the Astronomical Society of the Pacific, 115, 688 
\bibitem[Khokhlov(1991)]{1991A&A...245..114K} Khokhlov, A.~M.\ 1991, Astronomy and Astrophysics, 245, 114 
\bibitem[Krisciunas et al.(2017)]{2017AJ....154..211K} Krisciunas, K., et 
al.\ 2017, The Astronomical Journal, 154, 211 
\bibitem[Krisciunas et al.(2004)]{2004ApJ...602L..81K} Krisciunas, K., 
Phillips, M.~M., 
\& Suntzeff, N.~B.\ 2004, The Astrophysical Journal, 602, L81 
\bibitem[Law et al.(2009)]{2009PASP..121.1395L} Law, N.~M., et al.\ 2009, 
Publications of the Astronomical Society of the Pacific, 121, 1395 
\bibitem[Li et al.(2003)]{2003PASP..115..453L} Li, W., et al.\ 2003, 
Publications of the Astronomical Society of the Pacific, 115, 453 
\bibitem[Maeda et al.(2014)]{2014ApJ...794...37M} Maeda, K., Kutsuna, M., 
\& Shigeyama, T.\ 2014, The Astrophysical Journal, 794, 37 
\bibitem[Maguire et al.(2018)]{2018MNRAS.477.3567M} Maguire, K., et al.\ 
2018, Monthly Notices of the Royal Astronomical Society, 477, 3567 
\bibitem[Mandel et al.(2011)]{2011ApJ...731..120M} Mandel, K.~S., Narayan, 
G., \& Kirshner, R.~P.\ 2011, The Astrophysical Journal, 731, 120 
\bibitem[Margutti et al.(2014)]{2014ApJ...780...21M} Margutti, R., et al.\ 
2014, The Astrophysical Journal, 780, 21 
\bibitem[Marietta et al.(2000)]{2000ApJS..128..615M} Marietta, E., Burrows, 
A., 
\& Fryxell, B.\ 2000, The Astrophysical Journal Supplement Series, 128, 615 
\bibitem[Marion et al.(2009)]{2009AJ....138..727M} Marion, G.~H., et al.\ 
2009, The Astronomical Journal, 138, 727 
\bibitem[Marion et al.(2015)]{2015ApJ...798...39M} Marion, G.~H., et al.\ 
2015, The Astrophysical Journal, 798, 39 
\bibitem[Mattila et 
al.(2005)]{2005A&A...443..649M} Mattila, S., et al.\ 2005, Astronomy and Astrophysics, 443, 649 
\bibitem[Milisavljevic et al.(2013)]{2013ApJ...770L..38M} Milisavljevic, 
D., et al.\ 2013, The Astrophysical Journal, 770, L38 
\bibitem[Milisavljevic et al.(2015)]{2015ApJ...799...51M} Milisavljevic, 
D., et al.\ 2015, The Astrophysical Journal, 799, 51 
\bibitem[Nomoto(1982)]{1982ApJ...253..798N} Nomoto, K.\ 1982, The 
Astrophysical Journal, 253, 798 
\bibitem[Nomoto et al.(1984)]{1984ApJ...286..644N} Nomoto, K., Thielemann, 
F.-K., \& Yokoi, K.\ 1984, The Astrophysical Journal, 286, 644 
\bibitem[Nugent et al.(1995)]{1995ApJ...455L.147N} Nugent, P., et al.\ 
1995, The Astrophysical Journal, 455, L147 
\bibitem[Oke 
\& Sandage(1968)]{1968ApJ...154...21O} Oke, J.~B., \& Sandage, A.\ 1968, The Astrophysical Journal, 154, 21 
\bibitem[Parrent et al.(2011)]{2011ApJ...732...30P} Parrent, J.~T., et al.\ 
2011, The Astrophysical Journal, 732, 30 
\bibitem[Penney 
\& Hoeflich(2014)]{2014ApJ...795...84P} Penney, R., \& Hoeflich, P.\ 2014, The Astrophysical Journal, 795, 84 
\bibitem[Perlmutter et al.(1999)]{1999ApJ...517..565P} Perlmutter, S., et 
al.\ 1999, The Astrophysical Journal, 517, 565 
\bibitem[Persson et al.(2013)]{2013PASP..125..654P} Persson, S.~E., et al.\ 
2013, Publications of the Astronomical Society of the Pacific, 125, 654 
\bibitem[Phillips(1993)]{1993ApJ...413L.105P} Phillips, M.~M.\ 1993, The 
Astrophysical Journal, 413, L105 
\bibitem[Phillips et al.(2007)]{2007PASP..119..360P} Phillips, M.~M., et 
al.\ 2007, Publications of the Astronomical Society of the Pacific, 119, 
360 
\bibitem[Phillips et al.(2018)]{P18} Phillips, M.~M., et al.\ 2018, 
Publications of the Astronomical Society of the Pacific, submitted
\bibitem[Piro 
\& Bildsten(2008)]{2008ApJ...673.1009P} Piro, A.~L., \& Bildsten, L.\ 2008, The Astrophysical Journal, 673, 1009 
\bibitem[Piro 
\& Chang(2008)]{2008ApJ...678.1158P} Piro, A.~L., \& Chang, P.\ 2008, The Astrophysical Journal, 678, 1158 
\bibitem[R{\"o}pke et 
al.(2006)]{2006A&A...453..203R} R{\"o}pke, F.~K., et al.\ 2006, Astronomy and Astrophysics, 453, 203 
\bibitem[Rayner et al.(2003)]{2003PASP..115..362R} Rayner, J.~T., et al.\ 
2003, Publications of the Astronomical Society of the Pacific, 115, 362 
\bibitem[Riess et al.(1998)]{1998AJ....116.1009R} Riess, A.~G., et al.\ 
1998, The Astronomical Journal, 116, 1009 
\bibitem[Ruiz-Lapuente et al.(1997)]{1997ASIC..486.....R} Ruiz-Lapuente, 
P., Canal, R., 
\& Isern, J.\ 1997, NATO Advanced Science Institutes (ASI) Series C, 486,  
\bibitem[Sand et al.(2016)]{2016ApJ...822L..16S} Sand, D.~J., et al.\ 2016, 
The Astrophysical Journal, 822, L16 
\bibitem[Sand et al.(2018)]{2018ApJ...863...24S} Sand, D.~J., et al.\ 2018, 
The Astrophysical Journal, 863, 24 
\bibitem[Scolnic et al.(2018)]{2018ApJ...859..101S} Scolnic, D.~M., et al.\ 
2018, The Astrophysical Journal, 859, 101 
\bibitem[Seitenzahl et 
al.(2013)]{2013A&A...559L...5S} Seitenzahl, I.~R., et al.\ 2013, Astronomy and Astrophysics, 559, L5 
\bibitem[Shappee et al.(2013)]{2013ApJ...762L...5S} Shappee, B.~J., et al.\ 
2013, The Astrophysical Journal, 762, L5 
\bibitem[Silverman et al.(2012)]{2012MNRAS.425.1789S} Silverman, J.~M., et 
al.\ 2012, Monthly Notices of the Royal Astronomical Society, 425, 1789 
\bibitem[Silverman 
\& Filippenko(2012)]{2012MNRAS.425.1917S} Silverman, J.~M., \& Filippenko, A.~V.\ 2012, Monthly Notices of the Royal Astronomical Society, 425, 1917 
\bibitem[Simcoe et al.(2013)]{2013PASP..125..270S} Simcoe, R.~A., et al.\ 
2013, Publications of the Astronomical Society of the Pacific, 125, 270 
\bibitem[Stanishev et 
al.(2018)]{2018A&A...615A..45S} Stanishev, V., et al.\ 2018, Astronomy and Astrophysics, 615, A45 
\bibitem[Stritzinger et al.(2011)]{2011AJ....142..156S} Stritzinger, M.~D., 
et al.\ 2011, The Astronomical Journal, 142, 156 
\bibitem[Stritzinger et 
al.(2015)]{2015A&A...573A...2S} Stritzinger, M.~D., et al.\ 2015, Astronomy and Astrophysics, 573, A2 
\bibitem[Stritzinger et 
al.(2018)]{2018A&A...609A.134S} Stritzinger, M.~D., et al.\ 2018, Astronomy and Astrophysics, 609, A134 
\bibitem[Suzuki et al.(2012)]{2012ApJ...746...85S} Suzuki, N., et al.\ 
2012, The Astrophysical Journal, 746, 85 
\bibitem[Taubenberger et al.(2011)]{2011MNRAS.412.2735T} Taubenberger, S., 
et al.\ 2011, Monthly Notices of the Royal Astronomical Society, 412, 2735 
\bibitem[Thielemann et 
al.(1986)]{1986A&A...158...17T} Thielemann, F.-K., Nomoto, K., \& Yokoi, K.\ 1986, Astronomy and Astrophysics, 158, 17 
\bibitem[Thomas et al.(2011)]{2011ApJ...743...27T} Thomas, R.~C., et al.\ 
2011, The Astrophysical Journal, 743, 27 
\bibitem[Thomas et al.(2011)]{2011PASP..123..237T} Thomas, R.~C., Nugent, 
P.~E., 
\& Meza, J.~C.\ 2011, Publications of the Astronomical Society of the Pacific, 123, 237 
\bibitem[Timmes et al.(2003)]{2003ApJ...590L..83T} Timmes, F.~X., Brown, 
E.~F., \& Truran, J.~W.\ 2003, The Astrophysical Journal, 590, L83 
\bibitem[Tomasella et al.(2016)]{2016MNRAS.459.1018T} Tomasella, L., et 
al.\ 2016, Monthly Notices of the Royal Astronomical Society, 459, 1018 
\bibitem[Vacca et al.(2003)]{2003PASP..115..389V} Vacca, W.~D., Cushing, 
M.~C., 
\& Rayner, J.~T.\ 2003, Publications of the Astronomical Society of the Pacific, 115, 389 
\bibitem[Valenti et al.(2015)]{2015MNRAS.448.2608V} Valenti, S., et al.\ 
2015, Monthly Notices of the Royal Astronomical Society, 448, 2608 
\bibitem[Webbink(1984)]{1984ApJ...277..355W} Webbink, R.~F.\ 1984, The 
Astrophysical Journal, 277, 355 
\bibitem[Wheeler et al.(1998)]{1998ApJ...496..908W} Wheeler, J.~C., et al.\ 
1998, The Astrophysical Journal, 496, 908 
\bibitem[Whelan 
\& Iben(1973)]{1973ApJ...186.1007W} Whelan, J., \& Iben, I., Jr.\ 1973, The Astrophysical Journal, 186, 1007 
\bibitem[Wilk et al.(2018)]{2018MNRAS.474.3187W} Wilk, K.~D., Hillier, 
D.~J., 
\& Dessart, L.\ 2018, Monthly Notices of the Royal Astronomical Society, 474, 3187 
\bibitem[Wood-Vasey et al.(2008)]{2008ApJ...689..377W} Wood-Vasey, W.~M., 
et al.\ 2008, The Astrophysical Journal, 689, 377 
\end{thebibliography}
\end{document}